\documentclass[a4paper,11pt]{article}
\usepackage[utf8x]{inputenc}
\usepackage[square, numbers, comma, sort&compress]{natbib} % Use the natbib reference package - read up on this to edit the reference style;  
\usepackage{slashed} % feynman slash
\usepackage{amsmath, amssymb} % feynman graphs

\usepackage{graphicx}

\graphicspath{{./Pictures/}} % Specifies the directory where pictures are stored

\newcommand{\im}{\mathrm{i}}

\newcommand{\oh}{\frac{1}{2}}

\newcommand{\ddq}{\oh \int \! \! \frac{d^3q}{(2\pi)^3}}
\newcommand{\dvq}{\int \! \! \frac{d^4q}{(2\pi)^4}}
\newcommand{\dq}{\int \! \! \frac{d^3q}{(2\pi)^3}}

\newcommand{\dVf}{\int \! \! \frac{d^4q}{(2\pi)^4} V(|\vec p - \vec q\,|)}

\newcommand{\Cp}{\cos \varphi_+(q)}
\newcommand{\Cm}{\cos \varphi_-(q)}
\newcommand{\Sp}{\sin \varphi_+(q)}
\newcommand{\Sm}{\sin \varphi_-(q)}

\newcommand{\Cpp}{\cos \varphi_+(p)}
\newcommand{\Cmp}{\cos \varphi_-(p)}
\newcommand{\Spp}{\sin \varphi_+(p)}
\newcommand{\Smp}{\sin \varphi_-(p)}

\newcommand{\pj}{P_J(\hat p \cdot \hat q)}
\newcommand{\pjp}{P_{J+1}(\hat p \cdot \hat q)}
\newcommand{\pjm}{P_{J-1}(\hat p \cdot \hat q)}

\newcommand{\om}{\omega_+(q)}

\newcommand{\1}{\,\chi_1(q)} \newcommand{\2}{\,\chi_2(q)}\newcommand{\3}{\,\chi_3(q)}\newcommand{\4}{\,\chi_4(q)}
\newcommand{\5}{\,\chi_5(q)} \newcommand{\6}{\,\chi_6(q)}\newcommand{\7}{\,\chi_7(q)}\newcommand{\8}{\,\chi_8(q)}

\makeatletter
\newcommand*{\tol}[1]{$\overline{\hbox{#1}}\m@th$}
\makeatother

\title{Effective Chiral Symmetry Restoration for Heavy-Light Mesons}
\author{V. K. Sazonov, G. Schaffernak, R. F. Wagenbrunn}

\begin{document}

\maketitle

\begin{abstract}
We study the spectrum of heavy-light mesons within a model with linear instantaneous 
confining potential. The single-quark Green function and spontaneous breaking of chiral
symmetry are obtained from the Schwinger-Dyson (gap) equation. For the meson spectrum we derive
a Bethe-Salpeter equation (BSE).
We solve thiss equation numerically in the heavy-light limit and obtain effective 
restoration of chiral and $U(1)_A$ symmetries at large spins. 
\end{abstract}

\section{Introduction}
Confinement and chiral symmetry belong to the most important properties of low-energy QCD.
On the classical level the massless QCD Lagrangian is invariant under the 
$U(2)_L\times U(2)_R = SU(2)_R \times SU(2)_L \times U(1)_A \times U(1)_V$
group of chiral symmetry \cite{IntrCSKoch}.
However,  the $SU(2)_A$ and $U(1)_A$ parts of chiral symmetry are dynamically broken.
The $U(1)_A$ symmetry is also broken by a quantum anomaly \cite{AdlerU1, BJU1, Fujikawa}.

Dynamical symmetry breaking is responsible for the mass generation of low-lying hadrons.
For instance, from the chiral symmetry point of view the pion is a pseudo Goldstone boson associated
with the spontaneous broken axial part of chiral symmetry \cite{GellMann1960}.
On the other hand, recent lattice simulations demonstrated the existence of hadrons, even when
chiral symmetry is artificially restored \cite{Mario1, Mario2, Mario3}. This implies that not only the breaking of chiral
symmetry contributes to the hadron masses.
At the same time excited hadrons can be arranged in the approximate multiplets of chiral and $U(1)_A$ groups
\cite{G1, Cohen2002, Cohen2, G2, G3, G3p2, G4} and this may be considered as an indication of effective
restoration of chiral symmetry \cite{GlozmanRep}. However, such effective restoration still requires further experimental 
confirmation, which involves a discovery of missing hadronic states.

In case of effective restoration the system doesn't undergo a phase transition and
the chiral order parameter doesn't vanish. This means that the quark condensate of the vacuum
still persists, but becomes unimportant.
This may be understood considering the quarks in the meson rest frame.
In hadrons with large spins the quarks have 
a little probability to be in a state with low momentum and thus they decouple from the quark condensate.

The phenomenon of effective chiral symmetry restoration has been illustrated in \cite{GlozNefRib} within the chirally
symmetric confined model \cite{Yaouanc1, Yaouanc2}. This effect has been proven analytically and  by the direct calculations of the meson spectra
in the heavy-light case for the quadratic potential \cite{KalashnikovaHeavyLight} and in the light-light case for the linear potential \cite{WG1}, \cite{Wagenbrunn}.
The effective restoration has also been discussed within different approaches in \cite{Afonin, Shifman, Cata, DeGrand, Swan1, Swan2, Cohen3}.
In this chapter we consider the system of two quarks with masses $m_1$ and $m_2$, confined by a linear potential. 
Such potential is observed in Coulomb gauge lattice simulations \cite{Nakagawa} and can be considered
as the most realistic one from the phenomenological point of view. In case of massless quarks the Hamiltonian
(\ref{H}) obeys $SU(2)_L \times SU(2)_R \times U(1)_A \times U(1)_V$ symmetry.

In the heavy quark limit $m_2\rightarrow\infty$, the spin and the isospin of the heavy quark 
and the total angular momentum $j_l$ of the light quark are separately conserved \cite{Isgur1989, Flynn1992, Bardeen1993, Bardeen2003}. 
Heavy quark symmetry, which unifies the heavy spin and heavy flavor symmetries, leads to  
independent dynamics of the light quark 
with respect to the isospin and spin of the heavy quark. All mesonic
states can be classified by quantum numbers of the light quark.
% 
% In our case only the heavy-spin symmetry is relevant, since there are only two flavors 
% of quarks the first with mass $m_1$ and the second with mass $m_2$.
% 
For the fixed total angular momentum $j_l \neq 0$, 
heavy spin symmetry implies the existence of degenerate states with total spin $J = j_l \pm\frac{1}{2}$.
In Fig. 1 we present a general view of the spectrum respecting a heavy spin symmetry.
\begin{figure}[ht]
 \centering
   \includegraphics[width=2.1in]{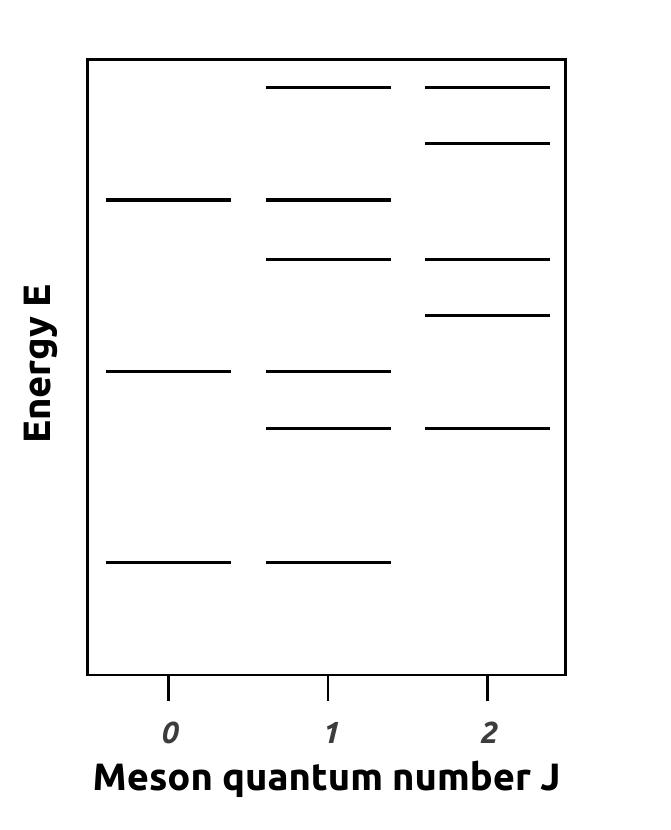}
   \rule{30em }{0.5pt}
   \caption{Expected spectrum for heavy spin symmetry}
   \label{fig:restoreheavy}
\end{figure}
The heavy flavor symmetry leads to a doubling of all states in the spectrum, because 
of the two possible orientations of the heavy quark isospin, i.e., each line in Fig. 1
should be considered as an isospin doublet.
The bare interaction in the model (\ref{H}) is isospin independent. It is also isospin independent 
in all orders of perturbation theory, since
we perform all computations in the large $N_c$ approximation, where all fermion loops are suppressed.
This leads to a doubling of all states in the spectrum also with respect to the light quark isospin projections.

In case of heavy-light system $m_1 = 0$, $m_2\rightarrow\infty$, at large angular momenta the restoration 
of chiral and $U(1)_A$ symmetries of the light quark is expected.
The chiral properties of mesons are defined by the representations of the parity-chiral group \cite{GlozmanRep},
consistent with left and right isospins $(I_L, I_R)$ and with $J^P$ quantum numbers. 
The isospin of the heavy quark doesn't influence the dynamics of the light quark, so the only
possible representation of the parity-chiral group for the heavy-light mesons is
\begin{equation}
  (I_L, I_R) \oplus (I_R, I_L) = (1/2, 0) \oplus (0, 1/2)\,,
\end{equation}
where $1/2$ is the isospin of the light quark, and the direct sum of two irreducible representations
forms a state of well defined parity.
The possible mesons for this representations are
\begin{equation}
 I_{light} = 1/2,\, J^P=(j_l\pm 1/2)^+ \,\longleftrightarrow \,\, I_{light} = 1/2,\, J^P=(j_l\pm 1/2)^-\,.
\end{equation}
The sign $\longleftrightarrow$ connects the states which must be degenerate in the chirally symmetric mode.
Therefore, the restoration of chiral symmetry implies the appearance of parity doublets. The general 
view of the meson spectrum with restored chiral symmetry is shown in Fig. 2.
\begin{figure}[ht]
 \centering
   \includegraphics[width=3in]{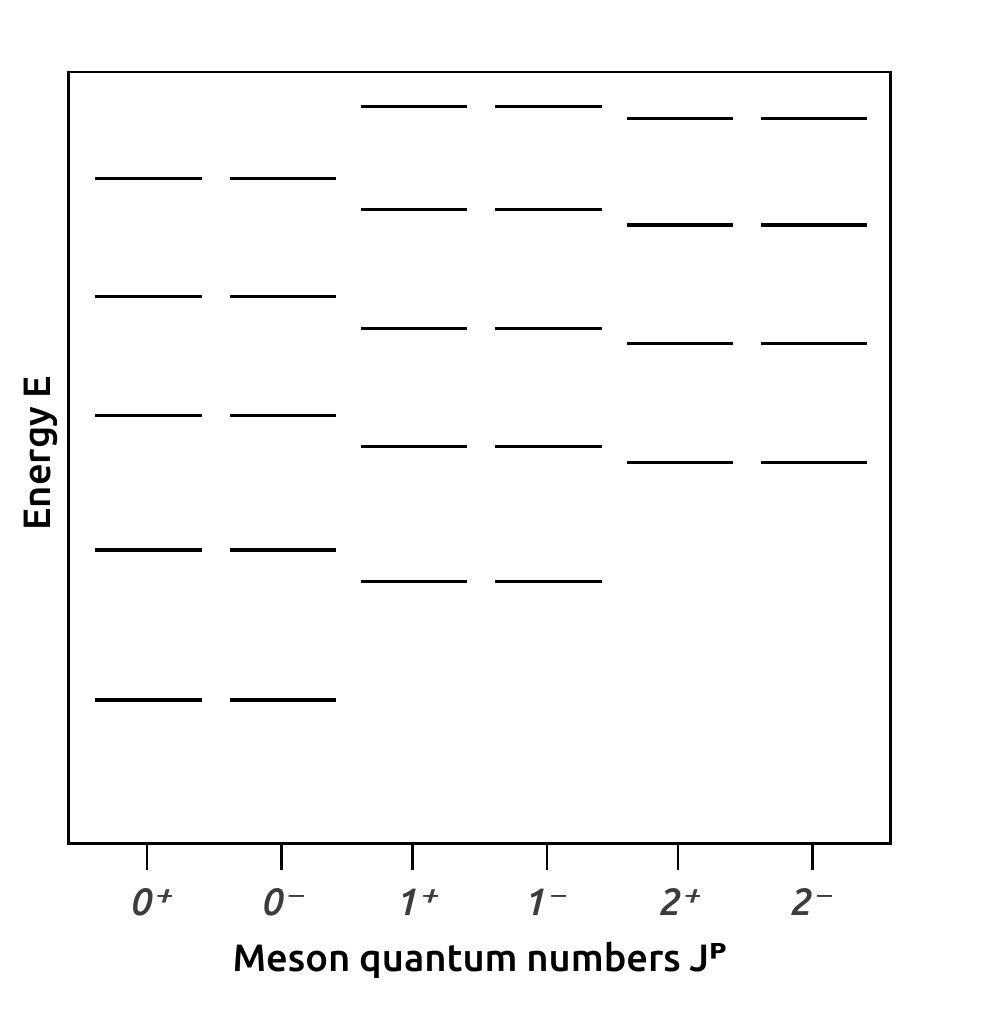}
   \rule{30em}{0.5pt}
   \caption{Expected spectrum given chiral symmetry restoration for the light quark}
   \label{fig:restorechiral}
\end{figure}
Combining the heavy spin and chiral symmetry one come to the expected spectrum for the heavy-light system in the chiral mode, Fig. 3.
\begin{figure}[ht]
 \centering
   \includegraphics[width=3.5in]{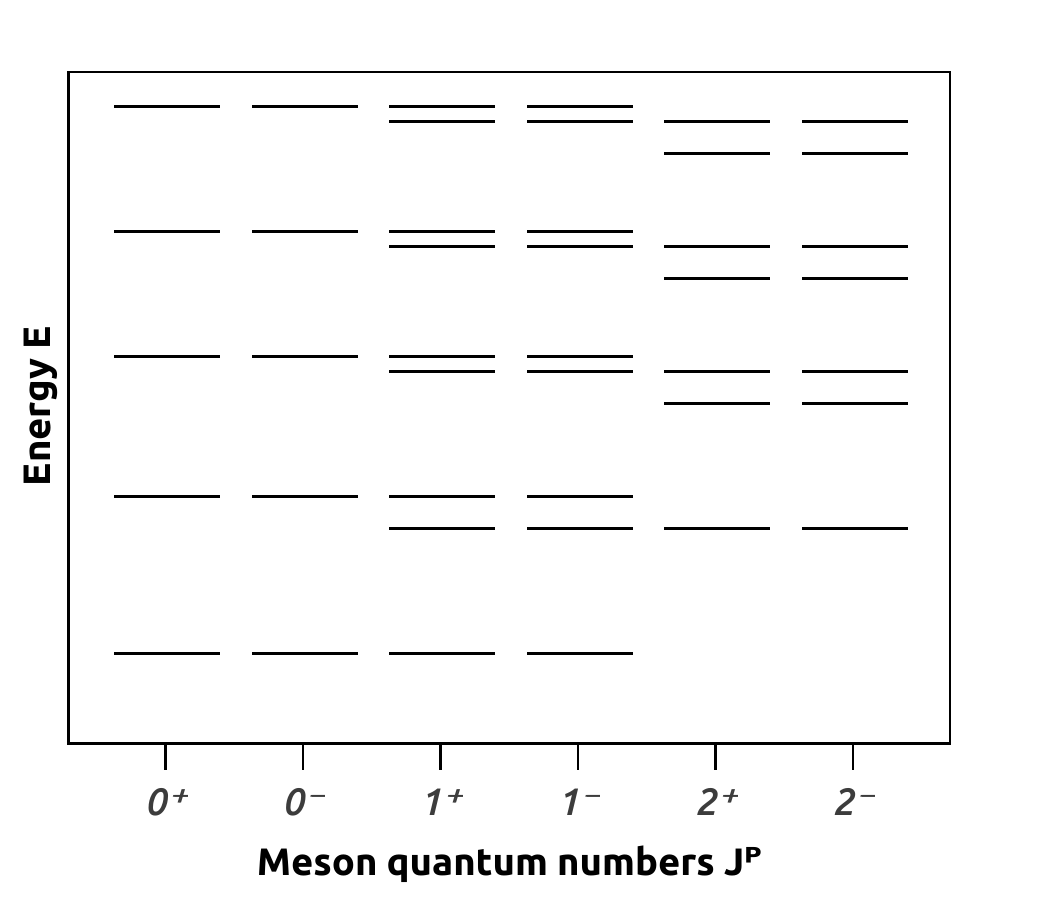}
   \rule{30em }{0.5pt}
   \caption{Expected spectrum for heavy spin symmetry and chiral symmetry}
   \label{fig:restoreboth}
\end{figure}

In the following sections within a framework of the chirally symmetric model with the linear confining potential
\cite{Yaouanc1, Yaouanc2} we study the aspects of effective restoration of chiral symmetry in heavy-light mesons.

The paper is organized as follows. In section \ref{sec:model} we describe
the model Hamiltonian and the confining potential. Section \ref{sec:ChSB} is dedicated
to chiral symmetry breaking in the vacuum in the considering model. We derive the general
Bethe-Salpeter equation for two quarks with different masses 
in the meson rest frame in section \ref{sec:BSE2m}.
In section \ref{sec:BSEhl}, taking the limit $m_2 \rightarrow \infty$, we derive BSE for the
heavy-light mesons and show the appearing of the heavy spin symmetry. In section \ref{sec:IRprop}
we demonstrate the cancellations of the infrared divergences in the Bethe-Salpeter equation.
Numerical results for the meson spectrum and for the corresponding wave functions are presented
in sections \ref{sec:NumRes} and \ref{sec:WF} respectively. In section \ref{sec:ERChS} we discuss
effective restoration of chiral symmetry and properties of the BSE without the spontaneous breaking
of chiral symmetry. We conclude in section \ref{sec:Concl}.

\section{Chirally symmetric model with confinement}
\label{sec:model}

To study effective restoration of chiral symmetry, one has to choose a confining model
with spontaneously broken chiral symmetry in $3+1$ dimensions. The latter remark is of great importance,
since effective chiral symmetry restoration doesn't appear in $1+1$ dimensions, because of the absence of 
rotational motion and spin.
Here we consider a model \cite{Yaouanc1, Yaouanc2} which may be viewed as a straight-forward generalization to the $3+1$ dimensional case
of the $1+1$ dimensional 't Hooft model \cite{tHooftM1, tHooftM2}. It was intensively studied in the context of chiral symmetry
breaking \cite{AdlerDavis} and chiral properties of hadronic states \cite{KalashnikovaHeavyLight, WG1, Shifman}. 

The Hamiltonian of the model is based on the instantaneous current-current interaction with linear potential of the Coulomb type
\begin{eqnarray} 
\hat{H} & = & \int d^3x\,\bar{\psi}_i(\vec{x},t)\left(-\im\,\vec{\gamma}\cdot
\vec{\bigtriangledown} + m_i \right)\psi_i(\vec{x},t) \nonumber \\
 &+& \frac12\int d^3
xd^3y\;J^a_\mu(\vec{x},t)K^{ab}_{\mu\nu}(\vec{x}-\vec{y})J^b_\nu(\vec{y},t)\,,
\label{H} 
\end{eqnarray} 
where $i = 1,2$ represents a summation over two different quark masses and quark fields $\psi_i = (u_i, d_i)$.
The quark current is $J_{\mu}^a(\vec{x},t)=\bar{\psi}_i(\vec{x},t)\gamma_\mu\frac{\lambda^a}{2}
\psi_i(\vec{x},t)$ and the potential looks like
\begin{equation} 
K^{ab}_{\mu\nu}(\vec{x}-\vec{y})=g_{\mu 0}g_{\nu 0}
\delta^{ab} V (|\vec{x}-\vec{y}|)\,,
\label{KK}
\end{equation}
where $a$, $b$ and $\mu$, $\nu$ are color and Lorentz indices respectively.

In case of two massless quarks the Hamiltonian obeys $SU(2)_L \times SU(2)_R \times U(1)_A \times U(1)_V$ symmetry
for both quarks.
When $m_1 = 0$ and $m_2  \neq 0$ the symmetry $SU(2)_L \times SU(2)_R \times U(1)_A \times U(1)_V$ is applicable
only to the first quark.
For further calculations we absorb the color Casimir factor in the string tension $\sigma$
\begin{equation}
\frac{\lambda^a \lambda^a}{4}V(r) = \sigma r\,.
\label{sigma_def}
\end{equation}

The Fourier transformation of the linear potential is ill-defined and requires an infrared regularization.
The regularization should be removed, or equivalently, the infrared limit should be taken in the final result. All 
physical observables must be finite in this limit and independent of the way,
how the potential was regularized. At the same time the single quark Green function can 
be divergent in the infrared limit, which demonstrates that a single quark cannot be observed within the 
framework of the considered and manifestly confined model.

There exist an infinite amount of physically equivalent regularizations. Following \cite{AlkoferAmundsen},
we define the potential as
\begin{equation}
V(p)= \frac{8\pi\sigma}{(p^2 + \mu_{\rm IR}^2)^2}\,.
\label{FV} 
\end{equation}

With the given prescription one can solve the Schwinger-Dyson (gap) and Bethe-Salpeter equations,
since all integrals in them are well defined. In case of the linear potential there are no ultraviolet divergences.
For other kinds of potentials the ultraviolet regularization and renormalization might be necessary.
For instance, this is the case if  the Coulomb interaction is added. In the following
we restrict ourselves only to the linear potential.

\section{Chiral symmetry breaking and the gap equation}
\label{sec:ChSB}

The interaction with gluon fields (or in our case self current-current interaction via confining potential)
leads to the dressing of the Dirac operator
\begin{equation}
D(p_0, \vec p\,) = \im S^{-1}(p_0, \vec p\,) = D_0(p_0, \vec p\,)- \Sigma(p_0, \vec p\,)\,,
\label{dressedDirac}
\end{equation}
where $\Sigma(p_0, \vec p\,)$ is the quark self energy and the bare Dirac operator is
\begin{equation}
D_0(p_0, \vec p\,) = p_0\gamma_0 - \vec p\cdot\vec\gamma - m\,.
\label{bareDirac}
\end{equation}
% The formula (\ref{dressedDirac}) is equivalent to the Schwinger-Dyson equation on Fig. \ref{fig:SchDyson}.
% 
% \begin{figure}[ht]
%  \centering
%    \includegraphics[width=5in]{./Figures/DS.pdf}
%    \rule{30em}{0.5pt}
%    \caption{Schwinger-Dyson equation in the diagrammatic representation.}
%    \label{fig:SchDyson}
% \end{figure}
% 
For an instantaneous interaction the self energy
\begin{equation}
  \im\,\Sigma(\vec p\,) = \dVf \gamma_0 \frac{1}{\im\, S_0^{-1}(q_0, \vec q\,) - \Sigma(\vec q\,)}\gamma_0
\label{selfE3d}
\end{equation}
is independent of the energy $p_0$. Representing the self energy in terms of Lorentz-invariant
amplitudes
\begin{equation}
\Sigma(\vec p\,) = A(p) - m + \vec \gamma \cdot \hat p (B(p) - p)
\label{SigmaAnsatz}
\end{equation}
and combining equations (\ref{dressedDirac}) and (\ref{selfE3d}) we arrive at a system
of two coupled integral equations
\begin{eqnarray}
\nonumber
A(p) =m + \frac{1}{2} \int \frac{d^3 q}{(2\pi)^3} V(|\vec p - \vec q\,|) \frac{A(q)}{\omega_q}\,,\\
\nonumber
B(p) = p + \frac{1}{2} \int \frac{d^3 q}{(2\pi)^3} V(|\vec p - \vec q\,|) \frac{B(q)}{\omega_q}\hat p \cdot\hat q\,,\\
\label{gapSysT0vac}
\end{eqnarray}
where 
\begin{equation}
\omega_p = \sqrt{A(p)^2 + B(p)^2}\,
\label{wp}
\end{equation}
is the single quark energy.
Introducing a Bogoliubov (chiral) angle
\begin{equation}
\sin\varphi_p \equiv \frac{A(p)}{\omega_p} \,,~~~~
\cos\varphi_p \equiv \frac{B(p)}{\omega_p} \,
\label{ChirAngDef}
\end{equation}
one can reduce the system (\ref{gapSysT0vac}) to a single integral equation
\begin{eqnarray}
p\sin\varphi_p - m \cos\varphi_p= \frac{1}{2}\! \int \!\!\frac{d^3 q}{(2\pi)^3} V(|\vec p - \vec q|) 
(\sin\varphi_q\cos\varphi_p - \hat p \!\cdot\!\hat q \sin\varphi_p\cos\varphi_q).
\label{gapeqVacchirA}
\end{eqnarray}
Then the expressions for the amplitudes $A(p)$ and $B(p)$ become
\begin{eqnarray}
A(p) = m + \frac{1}{2} \int \frac{d^3 q}{(2\pi)^3} V(|\vec p - \vec q\,|) \sin\varphi_q\,,\\
B(p) = p + \frac{1}{2} \int \frac{d^3 q}{(2\pi)^3} V(|\vec p - \vec q\,|) \hat p \cdot\hat q \cos\varphi_q
\label{ABchirA}
\end{eqnarray}
and the single quark energy can be determined via
\begin{eqnarray}
\omega_p = p\cos\varphi_p + m \sin\varphi_p + \frac{1}{2} \int \frac{d^3 q}{(2\pi)^3} V(|\vec p - \vec q|) 
(\sin\varphi_q\sin\varphi_p + \hat p \cdot\hat q~ \cos\varphi_p\cos\varphi_q)\,.
\label{SelfEchirA}
\end{eqnarray}
The single quark energy
\begin{eqnarray}
\omega_p = \frac{\sigma}{2 \mu_{IR}} + \omega_p^f\,
\label{wpdiv}
\end{eqnarray}
as the energy of any color state is infinite in the infrared limit. The
energies of all color-singlet states are finite and well defined. This is the manifestation of confinement
within the considered model.

At the same time, the integral in (\ref{gapeqVacchirA}) converges at $p = q$, since the infrared divergence
of the potential exactly cancels in the sum of the two integrand terms. Consequently, the chiral angle
and the dynamical mass
\begin{equation}
  M(p) = p \tan \varphi_p
\label{M(p)}
\end{equation}
are finite and can be found by the numerical solution of the gap equation (\ref{gapeqVacchirA}).
\begin{figure}[ht]
 \centering
   \includegraphics[width=3in]{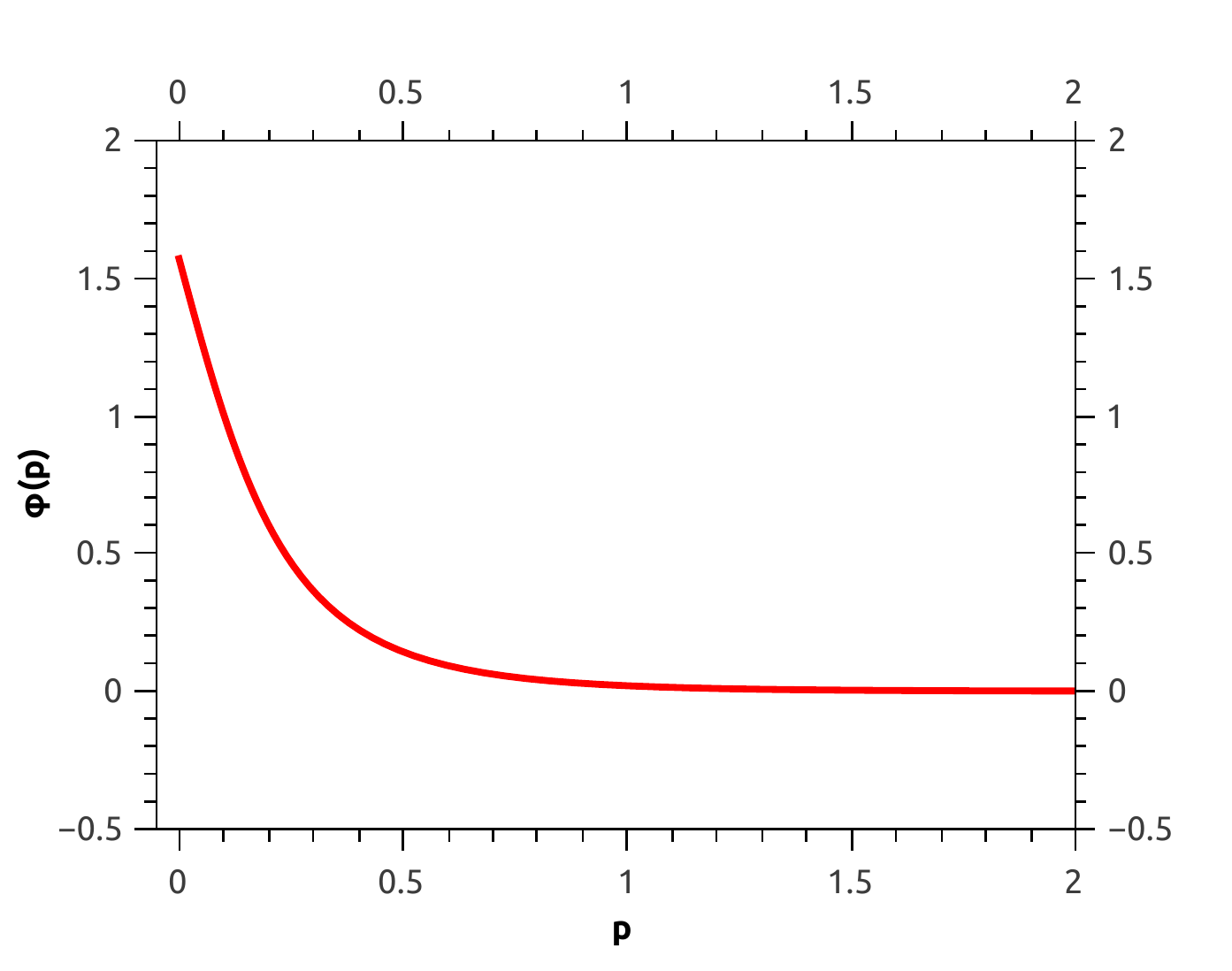}
   \rule{20em }{0.5pt}
   \caption{Chiral angle solution for m=0}
   \label{fig:chiralang}
\end{figure}
\begin{figure}[ht]
 \centering
   \includegraphics[width=3in]{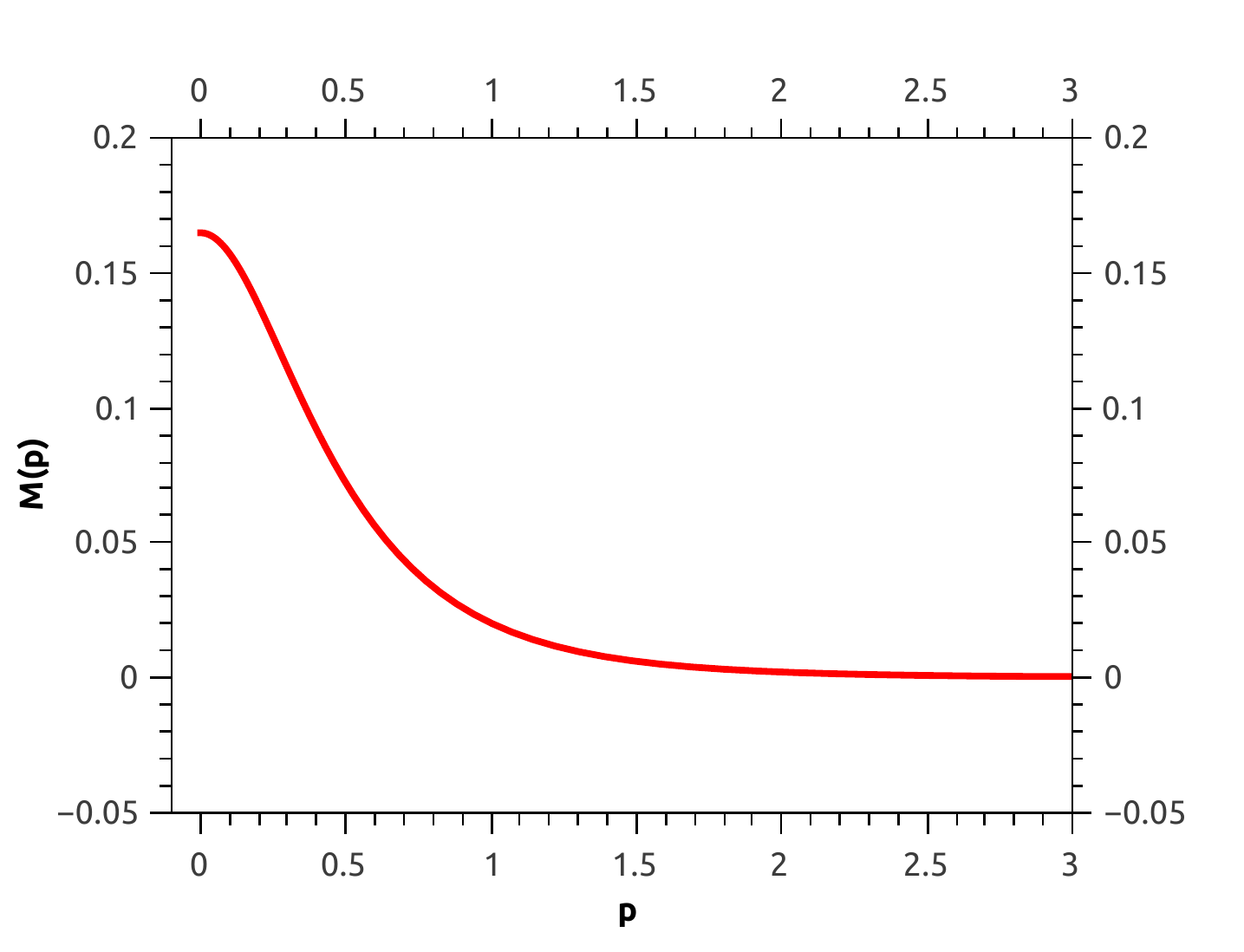}
   \rule{20em }{0.5pt}
   \caption{Dynamical mass function, momentum in units of $\sqrt{\sigma}$}
   \label{fig:dynmass}
\end{figure}
This solution leads to the non-zero quark condensate 
\begin{equation}
  <\bar q q> = -\frac{N_c}{\pi^2} \int_0^\infty dp~p^2\sin \varphi_p = -0.012\, \sigma^{3/2}\,,
\label{qcond}
\end{equation}
which signals spontaneous breaking of chiral symmetry.
As seen from the Fig. \ref{fig:chiralang} and Fig. \ref{fig:dynmass}, the effect of chiral symmetry breaking is large at low
momenta.
At large momenta these functions both approach zero. This property of the chiral angle is crucial for the
understanding of effective chiral symmetry restoration.

\section{Bethe-Salpeter equation for two quarks with different masses}
\label{sec:BSE2m}
To describe a bound state of  quarks and antiquarks we use the homogeneous 
Bethe-Salpeter equation (BSE) in the meson rest frame
\begin{equation}\label{eqn:bsegen} 
\chi(P,p)= -\im \dvq \, K(P,p,q) \, S_1(q+\frac{P}{2}) \, \chi(P,q) \, S_2(q-\frac{P}{2})\,,
\end{equation}
where $P$ is the mesons total momentum, $p$ is the relative momentum and $q$ is the loop momentum 
that has to be integrated over, $S_1$ and $S_2$ are the propagators 
for the quark and anti-quark. $K(P,p,q)$ is the Bethe-Salpeter kernel
and finally $\chi$ is the meson vertex function. 
In the rest frame of the meson the total momentum becomes
\begin{equation}\label{eqn:restframe} 
P^\nu = (m, 0, 0, 0)\,,
\end{equation}
with m being the mass of the bound-state. 
For the model with instantaneous interaction, 
the amplitude is energy independent and in the ladder approximation the equation reads
\begin{equation}\label{eqn:bse} 
\chi(m,\vec{q}\,)= -\im \dvq \, V(|\vec p - \vec q \,|) \,
\gamma_0\, S_1(q_0+\frac{m}{2}, \vec{q}\, ) \, \chi(m,\vec{q}\,) \, S_2(q_0-\frac{m}{2}, \vec{q}\, ) \,\gamma_0\,.
\end{equation}

To solve the Bethe-Salpeter equation (\ref{eqn:bse}) we expand the meson vertex function $\chi(m,\vec{q}\,)_{JM}^P$ into a set
of all possible independent Poincaré invariant amplitudes consistent with $J^P$ quantum numbers (see appendix A).
The mesons constructed from two quarks with different masses can be grouped in two 
categories:
\[
\text{Category A: }
\begin{cases}
J^{-}, J = 2n \\ 
J^{+}, J = 2n+1
\end{cases}
\text{Category B: }
\begin{cases}
J^{+}, J = 2n \\ 
J^{-}, J = 2n+1\,.
\end{cases}
\]
In appendix B we rewrite the BSE as systems of coupled integral equations, where each system corresponds to a
certain category of mesons. In terms of the wave function (see appendix C) the Bethe-Salpeter equation for
the category A and $J > 0$ is given as
\footnotesize
\begin{equation}
\nonumber
 \begin{split}
  (2 \omega_+(p) \mp &m) \, \psi_{1,\pm}(p) = \ddq V(k) \Big\{ \big[ (\, \Cmp \Cm + \Spp \Sp ) \pj \\
  &+ (  \Smp \Sm + \Cpp \Cp  ) \frac{(J+1) \pjp + J \pjm}{2J+1} \big] \, \psi_{1,\pm}(q)\\
  + &\big[( \Cmp \Cm - \Spp  \Sp ) \pj \hspace{1cm}\\
  &+ ( \Smp \Sm - \Cpp \Cp) \frac{(J+1) \pjp + J \pjm}{2J+1}\big] \, \psi_{1,\mp}(q)\\
  -& ( \Smp \Cp + \Cpp \Sm) \frac{\sqrt{J ( J+1)} }{2J+1} \big[ \pjp - \pjm \big] \, \psi_{2,\pm}(q)\\
  -& ( \Smp \Cp - \Cpp \Sm) \frac{\sqrt{J ( J+1)} }{2J+1} \big[ \pjp - \pjm \big] \, \psi_{2,\mp}(q)
  \Big\}\,,
 \end{split}
\label{cat1p1}
\end{equation}
\begin{equation}
 \begin{split}
  (2 \omega_+(p) \mp &m) \, \psi_{2,\pm}(p) = \ddq V(k) \Big\{ \big[ (\, \Cmp \Cm + \Spp \Sp ) \pj \\
  &+ (  \Smp \Sm + \Cpp \Cp  ) \frac{J \pjp + (J+1) \pjm}{2J+1} \big] \, \psi_{2,\pm}(q)\\
  - &\big[( \Cmp \Cm - \Spp  \Sp ) \pj \hspace{1cm}\\
  &+ ( \Smp \Sm - \Cpp \Cp) \frac{J \pjp + (J+1) \pjm}{2J+1}\big] \, \psi_{2,\mp}(q)\\
  -& ( \Smp \Cp + \Cpp \Sm) \frac{\sqrt{J ( J+1)} }{2J+1} \big[ \pjp - \pjm \big] \, \psi_{1,\pm}(q)\\
  +& ( \Smp \Cp - \Cpp \Sm) \frac{\sqrt{J ( J+1)} }{2J+1} \big[ \pjp - \pjm \big] \, \psi_{1,\mp}(q)
  \Big\}
 \end{split}
\label{cat1p2}
\end{equation}
\normalsize
and for category B with $J > 0$ as
\footnotesize
\begin{equation}
\nonumber
 \begin{split}
  (2 \omega_+(p) \mp &m) \, \psi_{1,\pm}(p) = \ddq V(k) \Big\{ \big[ (\, \Smp \Sm + \Cpp \Cp ) \pj \\
  &+ (  \Cmp \Cm + \Spp \Sp  ) \frac{(J+1) \pjp + J \pjm}{2J+1} \big] \, \psi_{1,\pm}(q)\\
  - &\big[( \Smp \Sm - \Cpp  \Cp ) \pj \hspace{1cm}\\
  &+ ( \Cmp \Cm - \Spp \Sp) \frac{(J+1) \pjp + J \pjm}{2J+1}\big] \, \psi_{1,\mp}(q)\\
  +& ( \Cmp \Sp + \Spp \Cm) \frac{\sqrt{J ( J+1)} }{2J+1} \big[ \pjp - \pjm \big] \, \psi_{2,\pm}(q)\\
  -& ( \Cmp \Sp - \Spp \Cm) \frac{\sqrt{J ( J+1)} }{2J+1} \big[ \pjp - \pjm \big] \, \psi_{2,\mp}(q)
  \Big\}\,,
 \end{split}
\label{cat2p1}
\end{equation}
\begin{equation}
 \begin{split}
  (2 \omega_+(p) \mp &m) \, \psi_{2,\pm}(p) = \ddq V(k) \Big\{ \big[ (\, \Smp \Sm + \Cpp \Cp ) \pj \\
  &+ (  \Cmp \Cm + \Spp \Sp  ) \frac{J \pjp + (J+1) \pjm}{2J+1} \big] \, \psi_{2,\pm}(q)\\
  + &\big[( \Smp \Sm - \Cpp  \Cp ) \pj \hspace{1cm}\\
  &+ ( \Cmp \Cm - \Spp \Sp) \frac{J \pjp + (J+1) \pjm}{2J+1}\big] \, \psi_{2,\mp}(q)\\
  +& ( \Cmp \Sp + \Spp \Cm) \frac{\sqrt{J ( J+1)} }{2J+1} \big[ \pjp - \pjm \big] \, \psi_{1,\pm}(q)\\
  +& ( \Cmp \Sp - \Spp \Cm) \frac{\sqrt{J ( J+1)} }{2J+1} \big[ \pjp - \pjm \big] \, \psi_{1,\mp}(q)
  \Big\}\,,
 \end{split}
\label{cat2p2}
\end{equation}
\normalsize
where the notations $k = |\vec p - \vec q\,|$, $\omega_+(p) = 
\oh (\omega_1(p)+\omega_2(p))$ and $\varphi_\pm (p) = \oh (\varphi_1(p)\pm \varphi_2(p))$
were used and $P_J(x)$ are Legendre polynomials.

For the case $J = 0$ the meson vertex function can be expanded in smaller number of basis elements than for $J > 0$
and the corresponding equations are shorter.
Then for category A the BSE reads as
\footnotesize
\begin{equation}
 \begin{split}
  (2 \omega_+(p) \mp m) \, \psi_{1,\pm}(p) = \ddq &V(k) \Big\{ \big[ (\, \Cmp \Cm + \Spp \Sp )  \\
  &+ (  \Smp \Sm + \Cpp \Cp  ) \, \hat p \cdot \hat q \, \big] \, \psi_{1,\pm}(q)\\
  + &\big[( \Cmp \Cm - \Spp  \Sp )  \hspace{1cm}\\
  &+ ( \Smp \Sm - \Cpp \Cp) \, \hat p \cdot \hat q \,\big] \, \psi_{1,\mp}(q)
  \Big\}
 \end{split}
\label{cat1J0}
\end{equation}
\normalsize
and for category B it is 
\footnotesize
\begin{equation}
\begin{split}
  (2 \omega_+(p) \mp m) \, \psi_{2,\pm}(p) = \ddq &V(k) \Big\{ \big[ (\, \Cpp \Cp + \Smp \Sm )  \\
  &+ (  \Spp \Sp + \Cmp \Cm  ) \, \hat p \cdot \hat q \, \big] \, \psi_{2,\pm}(q)\\
  + &\big[( \Cpp \Cp - \Smp  \Sm )  \hspace{1cm}\\
  &+ ( \Spp \Sp - \Cmp \Cm) \, \hat p \cdot \hat q \,\big] \, \psi_{2,\mp}(q)
  \Big\}
 \end{split}
\label{cat2J0}
\end{equation}
\normalsize

For each value of $J$ the categories A and B contain chiral partners, hence when chiral symmetry
is effectively restored, the equations (\ref{cat1p2}) and (\ref{cat2p2}) must coincide and the equations
(\ref{cat1J0}) and (\ref{cat2J0}) must be also identical.

\section{Bethe-Salpeter equation for the heavy-light system}
\label{sec:BSEhl}

To derive the Bethte-Salpeter equations for the heavy-light mesons we assume $m_1 = 0$ and $m_2 \rightarrow \infty$.
The latter limit means that the heavy quark chiral angle becomes constant $\varphi_2 \rightarrow \frac{\pi}{2}$.
To shorten notations we denote the light quark chiral angle $\varphi_1$ as $\varphi$.
The mass of the whole meson can be splitted into the mass of the heavy quark and the binding energy $\epsilon= m-m_2$. 
The heavy quark energy is replaced by $\omega_2(p)\rightarrow m_2 + \frac{\sigma}{2\mu_{IR}}$, 
where the last term is the corresponding infrared divergence. 
All wave functions, labeled by '-', propagating backwards in time, vanish in the heavy-light limit, when the interaction is instantaneous \cite{KalashnikovaHeavyLight}.
The remaining wave functions for the forward motion in time are identified as $\psi_{i,+} \equiv \psi_{i}$.

Equations for the category A for $J > 0$ become
\begin{equation}\label{eqn:hlbsewave1}
\nonumber
\begin{split}
\scalebox{0.9}{$\displaystyle\big( \omega(p) + \frac{\sigma}{2 \mu_{IR}} - \epsilon \big) \psi_1(p) = \ddq V(k)  
\bigg\{ \Big[ \sqrt{(1+\sin \varphi(p)) (1+\sin \varphi(q))}P_J(\hat p \cdot \hat q) $} &\\
\hspace*{1cm}\scalebox{0.9}{$\displaystyle+ \sqrt{\left(1-\sin \varphi(p)\right) (1-\sin \varphi(q))} \frac{(J+1)P_{J+1}(\hat p \cdot \hat q) + 
   J P_{J-1}(\hat p \cdot \hat q)}{2J +1} \Big]  \psi_1(q) $}&\\
\hspace*{1cm}\scalebox{0.9}{$\displaystyle+ \sqrt{\left(1-\sin \varphi(p)\right) (1-\sin \varphi(q))} 
\frac{\sqrt{J(J+1)}}{2J+1} \left[ P_{J+1}(\hat p \cdot \hat q)-P_{J-1}(\hat p \cdot \hat q) \right] \psi_2(q)\bigg\} $}\,,
\end{split}
\end{equation}
\begin{equation}\label{eqn:hlbsewave2}
\begin{split}
\scalebox{0.9}{$\displaystyle\big( \omega(p) + \frac{\sigma}{2 \mu_{IR}} - \epsilon \big) \psi_2(p) = \ddq V(k)  
\bigg\{ \Big[ \sqrt{(1+\sin \varphi(p)) (1+\sin \varphi(q))}P_J(\hat p \cdot \hat q) $} &\\
\hspace*{1cm}\scalebox{0.9}{$\displaystyle+ \sqrt{\left(1-\sin \varphi(p)\right) (1-\sin \varphi(q))} \frac{J P_{J+1}(\hat p \cdot \hat q) + 
   (J+1) P_{J-1}(\hat p \cdot \hat q)}{2J +1} \Big]  \psi_2(q) $}&\\
\hspace*{1cm}\scalebox{0.9}{$\displaystyle+ \sqrt{\left(1-\sin \varphi(p)\right) (1-\sin \varphi(q))} 
\frac{\sqrt{J(J+1)}}{2J+1} \left[ P_{J+1}(\hat p \cdot \hat q)-P_{J-1}(\hat p \cdot \hat q) \right] \psi_1(q)\bigg\} $} \hspace{-0.2cm}&
\end{split}
\end{equation}
and for $J=0$ there is only one equation
\begin{equation}\label{eqn:hlbsewavejz1}
\begin{split}
\scalebox{0.9}{$\displaystyle\big( \omega(p) + \frac{\sigma}{2 \mu_{IR}} - \epsilon \big) \psi_1(p) = \ddq V(k)  
 \Big[ \sqrt{(1+\sin \varphi(p)) (1+\sin \varphi(q))} $} &\\
\hspace*{1cm}\scalebox{0.9}{$\displaystyle+ \sqrt{\left(1-\sin \varphi(p)\right) (1-\sin \varphi(q))} \,\hat p \cdot \hat q \Big]  \psi_1(q)\,.  $}\hspace{-0.2cm}&
\end{split}
\end{equation}
Equations for category B for $J > 0$ become
\begin{equation}\label{eqn:hlbsewave21}
\begin{split}
\nonumber
\scalebox{0.9}{$\displaystyle\big( \omega(p) + \frac{\sigma}{2 \mu_{IR}} - \epsilon \big) \psi_1(p) = \ddq V(k)  
\bigg\{ \Big[ \sqrt{(1-\sin \varphi(p)) (1-\sin \varphi(q))}P_J(\hat p \cdot \hat q) $} &\\
\hspace*{1cm}\scalebox{0.9}{$\displaystyle+ \sqrt{\left(1+\sin \varphi(p)\right) (1+\sin \varphi(q))} \frac{J P_{J+1}(\hat p \cdot \hat q) + 
   (J+1) P_{J-1}(\hat p \cdot \hat q)}{2J +1} \Big]  \psi_1(q) $}&\\
\hspace*{1cm}\scalebox{0.9}{$\displaystyle+ \sqrt{\left(1+\sin \varphi(p)\right) (1+\sin \varphi(q))} 
\frac{\sqrt{J(J+1)}}{2J+1} \left[ P_{J+1}(\hat p \cdot \hat q)-P_{J-1}(\hat p \cdot \hat q) \right] \psi_2(q)\bigg\} $}\,,
\end{split}
\end{equation}
\begin{equation}\label{eqn:hlbsewave22}
\begin{split}
\scalebox{0.9}{$\displaystyle\big( \omega(p) + \frac{\sigma}{2 \mu_{IR}} - \epsilon \big) \psi_2(p) = \ddq V(k)  
\bigg\{ \Big[ \sqrt{(1-\sin \varphi(p)) (1-\sin \varphi(q))}P_J(\hat p \cdot \hat q) $} &\\
\hspace*{1cm}\scalebox{0.9}{$\displaystyle+ \sqrt{\left(1+\sin \varphi(p)\right) (1+\sin \varphi(q))} \frac{(J+1)P_{J+1}(\hat p \cdot \hat q) + 
   J P_{J-1}(\hat p \cdot \hat q)}{2J +1} \Big]  \psi_2(q) $}&\\
\hspace*{1cm}\scalebox{0.9}{$\displaystyle+ \sqrt{\left(1+\sin \varphi(p)\right) (1+\sin \varphi(q))} 
\frac{\sqrt{J(J+1)}}{2J+1} \left[ P_{J+1}(\hat p \cdot \hat q)-P_{J-1}(\hat p \cdot \hat q) \right] \psi_1(q)\bigg\} $}
\end{split}
\end{equation}
and for $J=0$ the equation is
\begin{equation}\label{eqn:hlbsewavejz2}
\begin{split}
\scalebox{0.9}{$\displaystyle\big( \omega(p) + \frac{\sigma}{2 \mu_{IR}} - \epsilon \big) \psi_2(p) = \ddq V(k)  
 \Big[ \sqrt{(1-\sin \varphi(p)) (1-\sin \varphi(q))} $} &\\
\hspace*{1cm}\scalebox{0.9}{$\displaystyle+ \sqrt{\left(1+\sin \varphi(p)\right) (1+\sin \varphi(q))} \,\hat p \cdot \hat q \Big]  \psi_2(q)\,.  $}\hspace{-0.2cm}&
\end{split}
\end{equation}

The given heavy-light equations exhibit heavy-quark spin symmetry, which involves a degeneration of states with
fixed parity and $J = j_l \pm \frac{1}{2}$. To prove heavy spin symmetry, we diagonalize the equations
(\ref{eqn:hlbsewave1} - \ref{eqn:hlbsewavejz2}), using the short-hand notations
\begin{equation}
\begin{split}
 r_+ &= \sqrt{(1+\sin \varphi(p)) (1+\sin \varphi(q))}\\
 r_- &= \sqrt{(1-\sin \varphi(p)) (1-\sin \varphi(q))} \\
 \int_q &:= \ddq V(k)\\
 \vec{\psi}(p) &= (\psi_1(p), \psi_2(p))
 \end{split}
\label{diagnot1}
\end{equation}
Symbolically, the matrix equations for categories A and B read 
\begin{equation}
C\,\vec{\psi}(p) = D\,\vec{\psi}(q)
\label{diagnot2}
\end{equation}
with $C$ already being in diagonal form for both categories.
\begin{equation}
C = C_A = C_{B} = 
\Bigg(
\begin{smallmatrix} \omega(p) + \frac{\sigma}{2 \mu_{IR}} - \epsilon & 0\\
0 & \omega(p) + \frac{\sigma}{2 \mu_{IR}} - \epsilon 
\end{smallmatrix} 
\Bigg)
\label{diagnot3}
\end{equation}
Matrix $D$ is different for categories A and B and given by
\begin{equation}
 \begin{split}
 \hspace*{-0.5cm} D_A = \Bigg(\begin{smallmatrix} \int_q \left(r_+ P_J \, +\, r_- \frac{(J+1)P_{J+1}(\hat p \cdot \hat q) + 
   J P_{J-1}(\hat p \cdot \hat q)}{2J +1}\right) & \int_q \left( r_-\frac{\sqrt{J(J+1)}}{2J+1} \left[ P_{J+1}(\hat p \cdot \hat q)-P_{J-1}(\hat p \cdot \hat q) 
   \right]\right) \\
  \int_q \left(r_-\frac{\sqrt{J(J+1)}}{2J+1} \left[ P_{J+1}(\hat p \cdot \hat q)-P_{J-1}(\hat p \cdot \hat q) \right]\right) & 
\int_q \left( r_+ P_J \, +\, r_- \frac{J P_{J+1}(\hat p \cdot \hat q) +  (J+1) P_{J-1}(\hat p \cdot \hat q)}{2J +1}\right)
  \end{smallmatrix}  \Bigg)\\[0.2cm]
 \hspace*{-0.4cm}   D_{B} = \Bigg(\begin{smallmatrix} \int_q \left(r_- P_J \, +\, r_+ \frac{J P_{J+1}(\hat p \cdot \hat q) + 
   (J+1) P_{J-1}(\hat p \cdot \hat q)}{2J +1}\right) & \int_q \left( r_+\frac{\sqrt{J(J+1)}}{2J+1} \left[ P_{J+1}(\hat p \cdot \hat q)-P_{J-1}(\hat p \cdot \hat q) 
   \right]\right) \\
  \int_q \left(r_+\frac{\sqrt{J(J+1)}}{2J+1} \left[ P_{J+1}(\hat p \cdot \hat q)-P_{J-1}(\hat p \cdot \hat q) \right]\right) & 
\int_q \left( r_- P_J \, +\, r_+ \frac{(J+1) P_{J+1}(\hat p \cdot \hat q) +  J P_{J-1}(\hat p \cdot \hat q)}{2J +1}\right)
  \end{smallmatrix}  \Bigg)\,.
 \end{split}
\label{diagnot4}
\end{equation}
$D_A$ and $D_B$ can be transformed into diagonal form
\begin{equation}
 \begin{split}
 \hspace*{-0.5cm}D_A = \Bigg(\begin{smallmatrix} \int_q \left( r_- P_{J-1} \, +\, r_+ P_J\right) & 0 \\
 0& \int_q \left(  r_- P_{J+1} \, +\, r_+ P_J\right)
  \end{smallmatrix}  \Bigg)\\[0.2cm]
 \hspace*{-0.4cm}   D_B = \Bigg(\begin{smallmatrix} \int_q \left( r_+ P_{J-1} \, +\, r_- P_J\right) & 0 \\
 0& \int_q \left(  r_+ P_{J+1} \, +\, r_- P_J\right)
  \end{smallmatrix}  \Bigg)\,.
 \end{split}
\label{diagEq}
\end{equation}
For the case $J = 0$ the matrices $C$ are
\begin{equation}
C_A^{J = 0}= C_{B}^{J = 0} = \omega(p) + \frac{\sigma}{2 \mu_{IR}} - \epsilon
\end{equation}
and the matrices $D$ are
\begin{equation}
 \begin{split}
 \hspace*{-0.5cm}D_A^{J = 0} = \int_q \left( r_+ P_{0} \, +\, r_- P_1\right)\\[0.2cm]
 \hspace*{-0.4cm}   D_B^{J = 0} = \int_q \left( r_+ P_{1} \, +\, r_- P_0\right)\,.
 \end{split}
\end{equation}

For each $J > 0$ the spectrum of $J^-$ consists of one part that coincides with $(J-1)^-$ and one part that 
coincides with $(J+1)^-$. The same is true for positive parity solutions, what finishes the proof
of the heavy quark spin symmetry.

\section{Infrared properties of the Bethe-Salpeter equation}
\label{sec:IRprop}

It is crucially to note that all infrared divergences, appearing in functions $A(p)$, 
$B(p)$ and $\omega(p)$ in the limit $\mu_{IR} \rightarrow 0$, exactly cancel in all
BS equations. This ensures the existence of finite solutions for the binding
energy spectrum and meson wave functions. Here we demonstrate the cancellation on the example of the $J = 0$
equation for the category A (\ref{eqn:hlbsewavejz1}).

Using the representation for the Dirac delta function
\begin{eqnarray}
  \lim_{\mu_{IR} \rightarrow 0} \frac{\mu_{IR}}{\pi^2}\,\int d^3q\,\frac{1}{((\vec p - \vec q)^2 + \mu^2_{IR})^2} f(\vec q)
  = \int d^3q\,\delta(\vec p - \vec q) f(\vec q) = f(\vec p)\,,
\end{eqnarray}
for $\mu_{IR} \rightarrow 0$ from the equation (\ref{eqn:hlbsewavejz1}) we obtain
\begin{equation}
\frac{\sigma}{\mu_{IR}} \psi_1(p) = \ddq\, 8\pi\sigma\, \frac{\pi^2\,\delta(\vec p - \vec q)}{\mu_{IR}}
 \Big[ (1+\sin \varphi(p)) + (1-\sin \varphi(p)) \,\hat p \cdot \hat p \Big]  \psi_1(p)\,.
\label{canc}
\end{equation}
The relation (\ref{canc}) is an identity and proofs the cancellation of infrared divergences in the BSE.

\section{Numerical results for the spectrum}
\label{sec:NumRes}

The diagonalized equations (\ref{diagEq}) may be solved numerically (see appendix D).
In Fig. \ref{fig:hlbind2} we present results for the binding energy spectrum of the heavy-light mesons.
Collecting the values for the heavy-quark spin multiplets, in table \ref{tab:hlbind3} we show 
explicitly the orbital and spin quantum numbers for the light quark $j_l = l + s$, as well as the corresponding 
heavy-spin multiplet for the meson. 
\begin{table}[!ht] 
\centering
\phantom{a}
\resizebox{\linewidth}{!}{\begin{tabular}{ c c | c c | c c | c c | c c }
  $0+\oh$    &    $1-\oh$  &    $2-\oh$  &   $1+\oh$   &  $2+\oh$    &   $3-\oh$   &  $4-\oh$    &   $3+\oh$   &  $4+\oh$    & $5-\oh$ \\[5pt] 
$\{0^-,1^-\}$&$\{0^+,1^+\}$&$\{1^-,2^-\}$&$\{1^+,2^+\}$&$\{2^-,3^-\}$&$\{2^+,3^+\}$&$\{3^-,4^-\}$&$\{3^+,4^+\}$&$\{4^-,5^-\}$&$\{4^+,5^+\}$ \\  \hline 
    1.84     &     2.04    &   2.84     &   2.82      &   3.47      &   3.48      &  4.02        &   4.02      &   4.49      &  4.49        \\
    2.87     &     3.08    &   3.67     &   3.63      &   4.18      &   4.19       &  4.65       &   4.65      &   5.07      &  5.07      \\
    3.71     &     3.91    &   4.38     &    4.32     &   4.80      &   4.82      &  5.23        &   5.22      &   5.60      &  5.60        \\
    4.42     &     4.61    &   5.01     &    4.94     &   5.37      &   5.39      &  5.76        &   5.75      &   6.09      &  6.10        
\end{tabular}}
\rule[-4mm]{8cm}{0.1mm} \vspace{-2.7mm}
\caption{orbital, $j_l$,  heavy-light meson binding energy}\label{tab:hlbind3}
\end{table}

The energy gap between opposite parity members of the same chiral multiplet, generated by chiral symmetry breaking, goes 
rapidly to zero with increasing angular momentum.
\begin{table}[!ht]
\centering
\phantom{a}
\begin{tabular}{ r |c| c|  c| c|  c}
spin multiplet&$\{0,1\}$&$\{1,2\}$&$\{2,3\}$&$\{3,4\}$&$\{4,5\}$ \\[0.1cm] \hline & & & & \\[-0.3cm] 
ground state gap&  0.20   &   0.02   &  0.01    &   0.00    &  0.00   \\   
\end{tabular}
\rule[-3mm]{8cm}{0.1mm} \vspace{-2.2mm}
\caption{energy gap between opposite parity members of the same chiral multiplet}\label{tab:hlgap}
\end{table}

In Fig. \ref{fig:Regge} the angular Regge trajectories for ground states of chiral partners are shown. 
The trajectories exhibit asymptotically linear behavior and coincide with each other.
These numerical results prove the effective chiral symmetry restoration for the heavy-light mesons at large angular momenta
for the considered model.
\newpage
\begin{figure}[ht]
 \centering
   \includegraphics[width=5.5in]{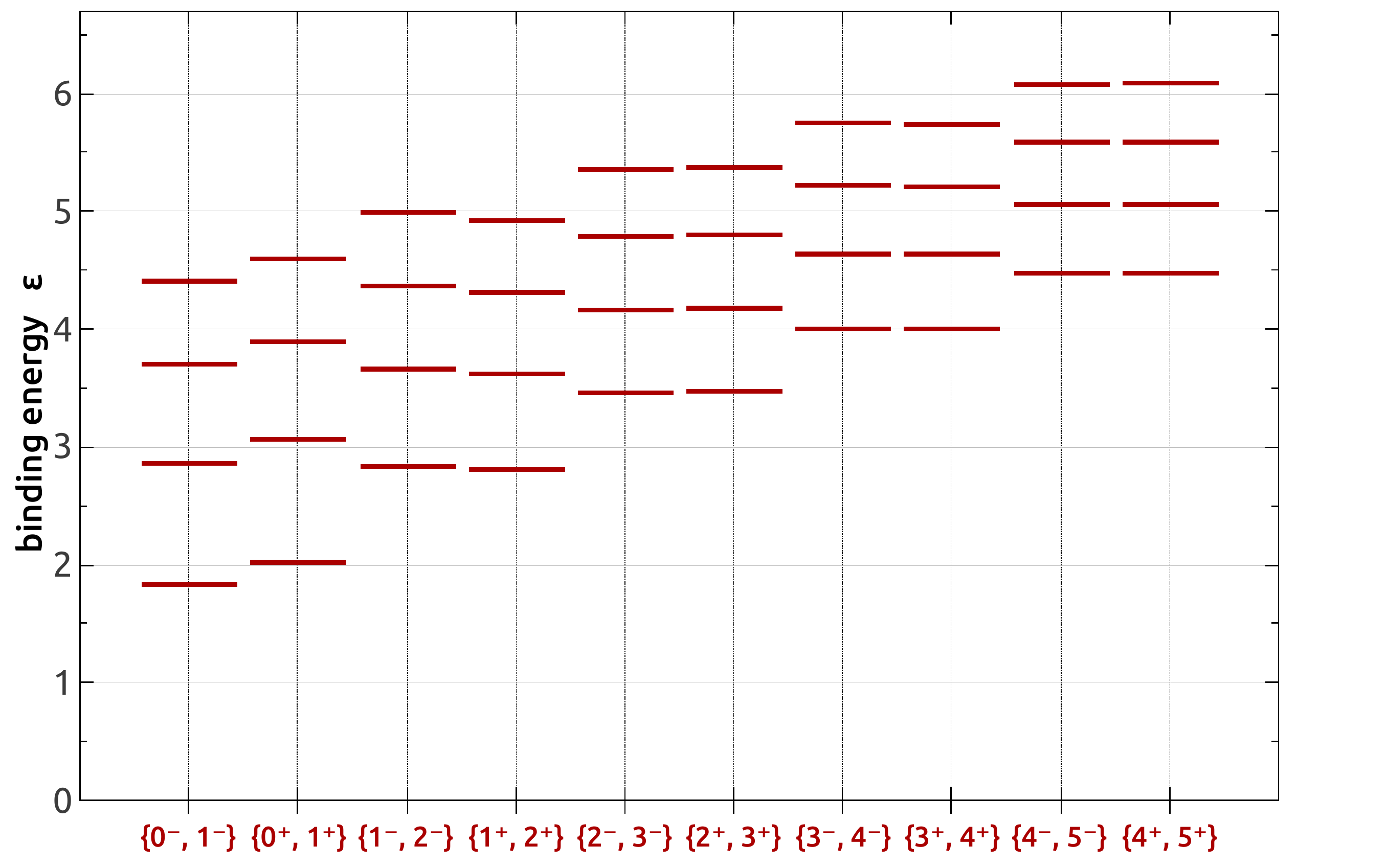}
   \rule{20em }{0.5pt}
   \caption{heavy-light meson binding energy in units of $\sqrt\sigma$}
   \label{fig:hlbind2}
\end{figure}
\begin{figure}[ht]
 \centering
   \includegraphics[width=3.5in]{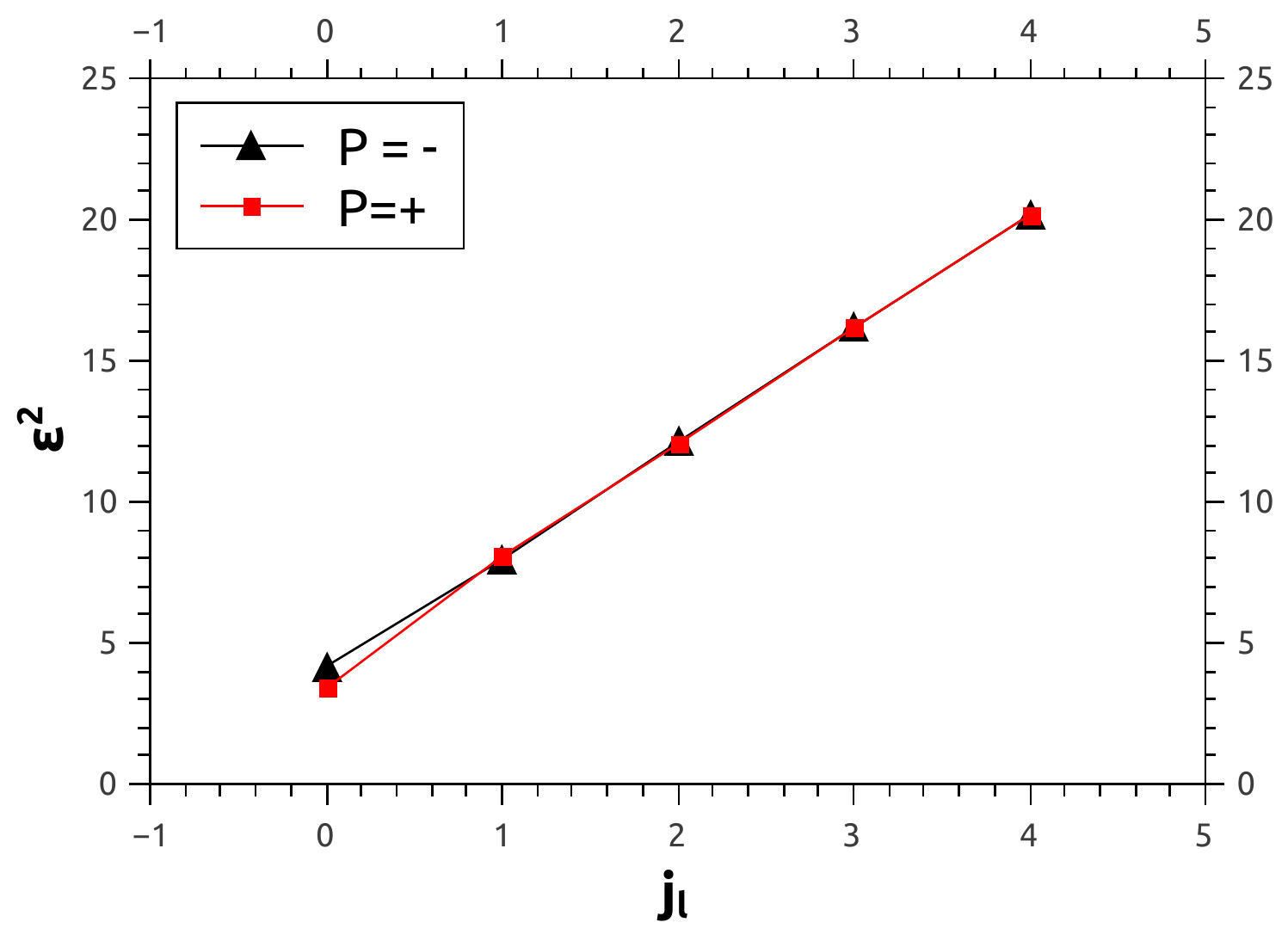}
   \rule{20em }{0.5pt}
   \caption{Angular Regge trajectories for ground states, $\varepsilon$ in units of $\sigma$}
   \label{fig:Regge}
\end{figure}
\newpage

\section{Wave functions}
\label{sec:WF}
\begin{figure}[ht]
 \centering
   \includegraphics[width=3.5in]{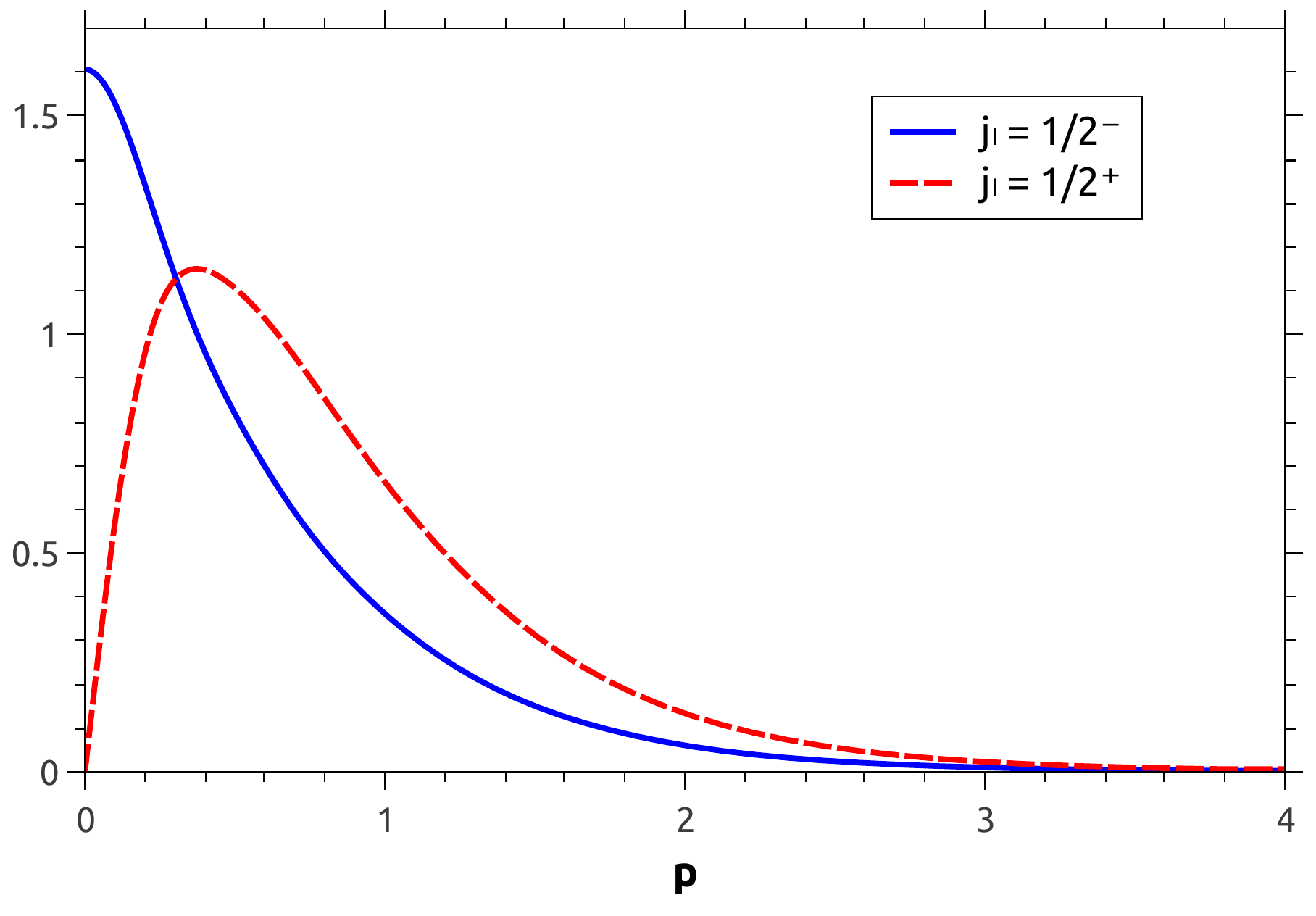}
   \rule{20em }{0.5pt}
   \caption{chiral partners for $j_l=\frac{1}{2}$, n=0}
   \label{fig:chiralpartner0}
\end{figure}

\begin{figure}[ht]
 \centering
   \includegraphics[width=3.5in]{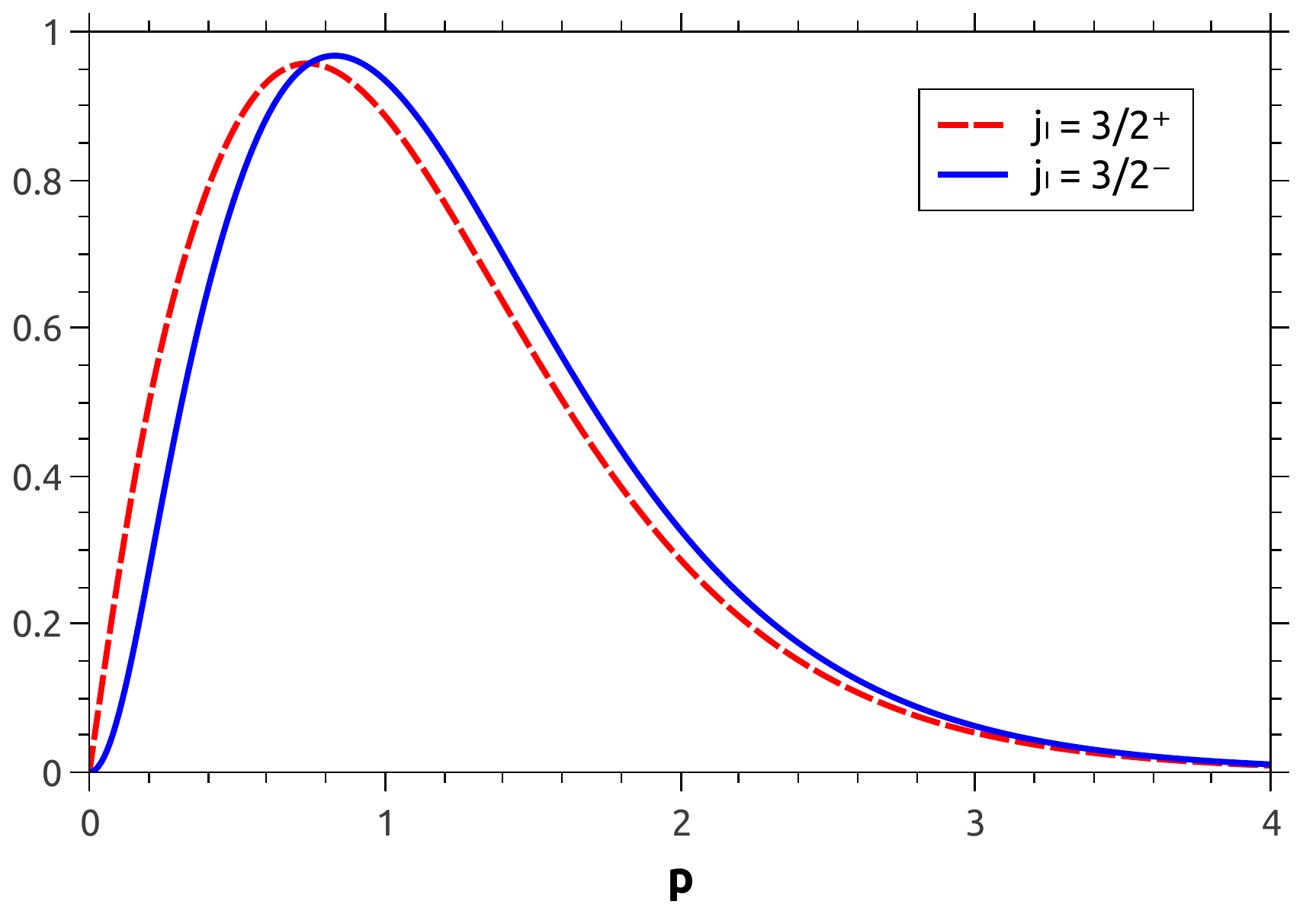}
   \rule{20em }{0.5pt}
   \caption{chiral partners for $j_l=\frac{3}{2}$, n=0}
   \label{fig:chiralpartner1}
\end{figure}

\begin{figure}[ht]
 \centering
   \includegraphics[width=3.5in]{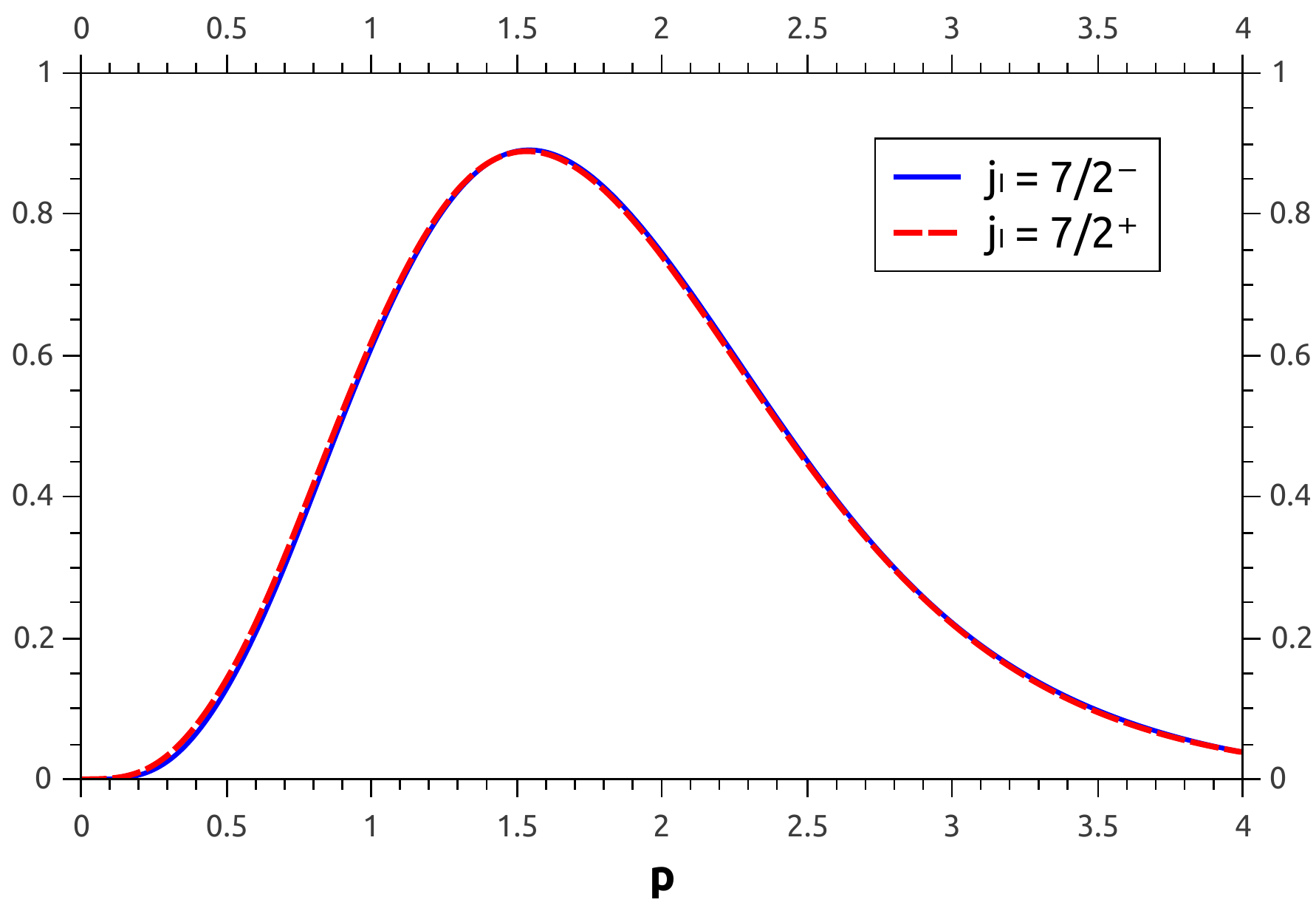}
   \rule{20em }{0.5pt}
   \caption{chiral partners for $j_l=\frac{7}{2}$, n=0}
   \label{fig:chiralpartner3}
\end{figure}

Since the spectrum exhibits the effective restoration of chiral symmetry, it must be seen also at the
level of wave functions. Continuing the classification of the states with respect to the total angular
momentum of the light quark, we compare the chiral partners' ground state wave functions for different values of $j_l$.
As can be seen from the respective plots, the ground state multiplets with $j_l = \frac{1}{2}$ have different wave function shapes,
while for $j_l = \frac{3}{2}$ the wave functions are much closer to each other and they practically coincide for $j_l = \frac{7}{2}$, see Figures 
\ref{fig:chiralpartner0}-\ref{fig:chiralpartner3}.
\clearpage
\clearpage

\section{Effective restoration of chiral symmetry for mesons}
\label{sec:ERChS}

In the previous sections we discussed numerical aspects of the effective restoration of chiral symmetry, here we perform some 
more analytical considerations. The increase of the light quark total angular momentum causes the grow of the 
quark relative momentum inside the meson. This can be also seen from the wave function plots, where the wave function's peaks
shift to the right with the increase of $j_l$. Then, effectively, for high $j_l$ only the high momenta contribute to the
integrals in the Bethe-Salpeter equation. The dynamical mass $M(p)$ and chiral
angle $\varphi(p)$ approach zero rather fast for $p \rightarrow \infty$ and effectively vanish in the BSE. 
Here we consider the consequences of such a vanishing.

In terms of notations (\ref{diagnot1}) - (\ref{diagnot4}) the
diagonalized version of equations (\ref{eqn:hlbsewave1}) - (\ref{eqn:hlbsewavejz2}) can be written as
\begin{equation}
 \begin{split}
 \hspace*{-0.5cm}\text{A:}~~~C \vec\psi(p) = \Bigg(\begin{smallmatrix} \int_q \left( r_- P_{J-1} \, +\, r_+ P_J\right) & 0 \\
 0& \int_q \left(  r_- P_{J+1} \, +\, r_+ P_J\right)
  \end{smallmatrix}  \Bigg)\vec\psi(p)\,,\\[0.2cm]
 \hspace*{-0.4cm}\text{B:}~~~   C\vec\psi(p) = \Bigg(\begin{smallmatrix} \int_q \left( r_+ P_{J-1} \, +\, r_- P_J\right) & 0 \\
 0& \int_q \left(  r_+ P_{J+1} \, +\, r_- P_J\right)
  \end{smallmatrix}  \Bigg)\vec\psi(p)\,,
 \end{split}
\label{diagEqAB}
\end{equation}
where the first and second equations describe categories A and B of mesons respectively.
When the chiral angle is set to zero, the coefficients $r_+$ and $r_-$ are $r_+(\varphi = 0) = 1$,\\ $r_-(\varphi = 0) = 1$,
therefore the equations for categories A and B from (\ref{diagEqAB}) become identical.
And the masses of the states with opposite parities must coincide, which is the direct signal of chiral restoration.

The analogous considerations are applicable to the case, 
where $m_1 = 0$ and $m_2$ is an arbitrary finite constant 
(equations (\ref{cat1p1})-(\ref{cat2J0})). Therefore, in such
kind of mesons the effective restoration of chiral symmetry is also expected.

\section{Conclusions}
\label{sec:Concl}
Within a model with the linear instantaneous Coulomb-like confining potential between quark currents
we have demonstrated a fast effective restoration of chiral symmetry in the spectrum of heavy-light mesons.
Effective chiral restoration leads to the degenerate masses of chiral partners (mesons with opposite parity)
and to the corresponding degeneration in the wave functions.

Chiral symmetry breaking appears in this model in the standard way through the non-perturbative 
quark self-interaction, generating a non-zero dynamical mass of the light quark. The dynamical mass is momentum-dependent
and vanishes at large momenta. For the heavy-light mesons with large $J$ the typical momentum of the light
quark is high, therefore it has a small effective dynamical mass and all quantum (loop) effects are suppressed
in this limit.

To describe bound states of the heavy-light system we derived the Bethe-Salpeter equation
for two quarks with different masses, which corresponds to a system of coupled integral equations. 
Then, by taking the limit $m_1 \rightarrow 0$, $m_2 \rightarrow \infty$ we obtained a heavy-light version of the equations
and proved the existence of heavy spin symmetry in this limit.
When the dynamical mass is zero this equations form exact chiral multiplets (parity doublets).
This is effective chiral restoration. 
Since the main physics is the same, the system of one quark with finite, not zero mass and 
one massless quark must also provide effective restoration of chiral symmetry.

\section{Acknowledgments}

VKS and GS acknowledge L. Ya. Glozman and M. Pak for helpful discussions.
This work was supported by the Austrian Science Fund (projects No. P21970-N16).

\section{Appendix A}

When the mesons are constructed, they can be classified with respect to $J^{PC}$
quantum numbers and fall into four categories \cite{Wagenbrunn}:
\[
\text{Category I: }
\begin{cases}
J^{-+}, J = 2n \\ %unsichtbare zeichen, benötigt für LEERSEITEN !!
J^{+-}, J = 2n+1
\end{cases}
\quad\text{Category II: }
\begin{cases}
J^{++}, J = 2n \\ %unsichtbare zeichen, benötigt für LEERSEITEN !!
J^{--}, J = 2n+1
\end{cases}
\]
\[
\text{Category III: }
\begin{cases}
J^{--}, J = 2n \\ %unsichtbare zeichen, benötigt für LEERSEITEN !!
J^{++}, J = 2n+1
\end{cases}
\quad\text{Category IV: }
\begin{cases}
J^{+-}, J = 2n \\ %unsichtbare zeichen, benötigt für LEERSEITEN !!
J^{-+}, J = 2n+1\,.
\end{cases}
\]
In case of mesons built from the quarks with different masses, the charge parity $C$ is not a well
defined quantum number. The classification of states should be done according only to the parity $P$
and to the angular momentum $J$. Then to form states with definite parity and undefined charge parity
the categories I and III mix and produce category A, the categories II and IV produce category B.
\[
\text{Category A: }
\begin{cases}
J^{-}, J = 2n \\ %unsichtbare zeichen, benötigt für LEERSEITEN !!
J^{+}, J = 2n+1
\end{cases}
\quad\text{Category B: }
\begin{cases}
J^{+}, J = 2n \\ %unsichtbare zeichen, benötigt für LEERSEITEN !!
J^{-}, J = 2n+1\,.
\end{cases}
\]
As it is pointed out in \cite{Wagenbrunn}, the instantaneous interaction in the Bethe-Salpeter
equation leads to the absence of the states of the category IV in the spectrum. However,
this doesn't affect the case with different quarks' masses and the category B should be still
considered as a union of categories II and IV.

The parameterizations for the vertex functions of categories A and B follow from their construction
and are the sums of corresponding parameterizations of categories I, III and II, IV.
Then using the formulas (A9), (A12), (A15), (A18) from \cite{Wagenbrunn} one can obtain the meson rest 
frame vertex functions for the instantaneous interaction.

The vertex function of the category A is
\footnotesize
\begin{eqnarray}
\nonumber
\chi^P_{JM}(m, \vec p) = \gamma_5 Y_{JM}(\hat p)\chi_1(p) 
+ m\gamma_0\gamma_5 Y_{JM}(\hat p)\chi_2(p)\\
\nonumber
+ m\gamma_0\gamma_5 \{Y_{J + 1}(\hat p)\otimes \vec\gamma\}_{JM} \chi_3(p)
+ m\gamma_0\gamma_5 \{Y_{J - 1}(\hat p)\otimes \vec\gamma\}_{JM} \chi_4(p)\\
\nonumber
+ m \{Y_{J}(\hat p)\otimes \vec\gamma\}_{JM} \chi_5(p) 
+ \gamma_5 \{Y_{J + 1}(\hat p)\otimes \vec\gamma\}_{JM} \chi_{6}(p)\\
+ \gamma_5 \{Y_{J - 1}(\hat p)\otimes \vec\gamma\}_{JM} \chi_{7}(p)
+ \gamma_0 \{Y_{J}(\hat p)\otimes \vec\gamma\}_{JM} \chi_{8}(p)\,,
\label{chibasis1XE}
\end{eqnarray}
\normalsize
where $\{\vec{a}_{J_1}\otimes \vec{b}_{J_2}\}_{JM}$ is the coupling of two spherical tensors of rank $J_1$ and $J_2$ to a spherical
tensor of rank $J$, $Y_{JM}(\hat p)$ are spherical harmonics and we denote $p = |\vec p|$.
For $J = 0$ the components $4,~5,~7,~8$ are absent.

The vertex function of the category B is
{\footnotesize
\begin{eqnarray}
\nonumber
\chi^P_{JM}(m, \vec p) = Y_{JM}(\hat p)\chi_1(p) 
+ \{Y_{J + 1}(\hat p)\otimes \vec\gamma\}_{JM} \chi_2(p)\\
\nonumber
+ \{Y_{J - 1}(\hat p)\otimes \vec\gamma\}_{JM} \chi_3(p)
+ m\gamma_5 \{Y_{J}(\hat p)\otimes \vec\gamma\}_{JM} \chi_4(p)\\
\nonumber
+ m \gamma_0 \{Y_{J + 1}(\hat p)\otimes \vec\gamma\}_{JM} \chi_5(p)
+ m \gamma_0 \{Y_{J - 1}(\hat p)\otimes \vec\gamma\}_{JM} \chi_6(p)\\
+ m \gamma_0 Y_{JM}(\hat p) \chi_7(p) 
+ \gamma_0 \gamma_5 \{Y_{J}(\hat p)\otimes \vec\gamma\}_{JM} \chi_{8}(p)\,.
\label{chibasis2XU}
\end{eqnarray}
}
\normalsize
For $J = 0$ the components $3,~4,~6,~8$ are absent.

\section{Appendix B}

Here we sketch the the derivation of the coupled integral Bethe-Salpeter equations for the 
category A, since the equations for the category B may be treated in the same way.

The vertex function in the rest frame doesn't depend on $p_0$ and it is easy
to take the $p_0$ integral in the Bethe-Salpeter equation (\ref{eqn:bse})
\begin{equation}\label{eqn:bse2}
\begin{split}
\hspace*{-0.1cm}\scalebox{0.83}{$\chi (m,\vec{p}\,) =  \displaystyle\dq V(k)\, \gamma_0 \, \Big\{
\dfrac{\gamma_0 \omega_1(q) + A_1(q) - \vec{\gamma}\cdot \hat{q}\, B_1(q) }{2 \omega_1(q)} 
\,  \chi (m,\vec{q}\,) \, \dfrac{\gamma_0(\omega_1(q)-m)+ A_2(q) - \vec{\gamma}\cdot \hat{q}\, B_2(q) }{(\omega_1(q)-m)^2-\omega_2^{\,2}(q)}$}& \\
\scalebox{0.83}{$ + \dfrac{\gamma_0 (\omega_2(q)+m) + A_1(q) - \vec{\gamma}\cdot \hat{q}\, B_1(q) }{(\omega_2(q)+m)^2-\omega_1^{\,2}(q)} 
\,  \chi (m,\vec{q}\,) \, \dfrac{\gamma_0 \omega_2(q)+ A_2(q) - \vec{\gamma}\cdot \hat{q}\, B_2(q) }{2 \omega_2(q)} 
 \Big\} \,\gamma_0$}\,. &
 \end{split}
\end{equation}
Then we substitute the vertex meson functions $\chi^P_{JM}(m, \vec p)$ from (\ref{chibasis1XE}) and (\ref{chibasis2XU}) to the BSE.
The interaction kernel mixes the initial Dirac structures. To match left hand side BSE Dirac structures
with right hand side ones, we project out the functions $\chi_i(p)$ on the left hand side. The projection
involves taking the traces of Dirac matrices and can be done by using of formulas (B1-B6) from \cite{Wagenbrunn}.
This leads to a system of eight coupled integral equations for the basis elements $\chi_i$. 
However, not all of these equations are linear independent. Introducing new functions
\footnotesize
\begin{equation}
 A_1(q) = \Cm \1 - \sqrt{\frac{J+1}{2J+1}} \Sm \6 + \sqrt{\frac{J}{2J+1}} \Sm \7\,,
\end{equation}
\begin{equation}
 A_2(q) = A_1(q) + 2 \om \left\{ \Sp \2 - \sqrt{\frac{J+1}{2J+1}} \Cm \3 + \sqrt{\frac{J}{2J+1}} \Cp \4 \right\}\,,
\end{equation}
\begin{equation}
 A_3(q) = \sqrt{\frac{J}{2J+1}} \Cp \6 + \sqrt{\frac{J+1}{2J+1}} \Cp \7 + \Sp \8\,,
\end{equation}
\begin{equation}
 A_4(q) = A_3(q) + 2 \om \left\{ \sqrt{\frac{J}{2J+1}} \Sm \3 + \sqrt{\frac{J+1}{2J+1}} \Sm \4 + \Cm \5 \right\}\,,
\end{equation}
\normalsize
we end up with four coupled integral equations 
\footnotesize
\begin{equation}
 \begin{split}
A_1(p)= \ddq V(k) \Bigg\{ \Big[\Cmp \Cm \pj \hspace*{6cm}&\\ 
+\Smp\Sm \frac{(J+1)\pjp+ J\pjm}{2J+1}\Big] \times \left[\frac{A_1(q)}{\om}+\frac{m^2 }{4\,\om}\frac{ A_2(q)}{\omega_+^{\,2}-\frac{m^2}{4}}\right]&\\
-\Smp \Cp\frac{\sqrt{J(J+1)}}{2J+1}\left[ \pjp - \pjm \right] \left[\frac{A_3(q)}{\om}+\frac{m^2 }{4\,\om}\frac{A_4(q)}{\omega_+^{\,2}-\frac{m^2}{4}}\right]&
\Bigg\}
 \end{split}
\end{equation}
\begin{equation}
 \begin{split}
A_2(p)= A_1(p) + \frac{\omega_+(p)}{2} \dq V(k) \Bigg\{ \Big[ \Spp \Sp \pj \hspace*{4cm}&\\ 
+\Cpp\Cp \frac{(J+1)\pjp+ J\pjm}{2J+1} \Big] \frac{ A_2(q)}{\omega_+^{\,2}-\frac{m^2}{4}}&  \\
-\Cpp \Sm \frac{\sqrt{J(J+1)}}{2J+1} \left[ \pjp - \pjm \right]\frac{ A_4(q)}{\omega_+^{\,2}-\frac{m^2}{4}}& \Bigg\}
 \end{split}
\end{equation}
\vspace{0.2cm}
\begin{equation}
 \begin{split}
A_3(p)= \ddq V(k) \Bigg\{ \Big[\Spp \Sp \pj\hspace*{6cm}&\\
+\Cpp \Cp \frac{J\pjp+ (J+1)\pjm}{2J+1}\Big] \times \left[\frac{A_3(q)}{\om}+\frac{m^2 }{4\,\om}\frac{ A_4(q)}{\omega_+^{\,2}-\frac{m^2}{4}}\right]&\\
-\Cpp\Sm\frac{\sqrt{J(J+1)}}{2J+1}\left[ \pjp - \pjm \right] \left[\frac{A_1(q)}{\om}+\frac{m^2 }{4\,\om}\frac{A_2(q)}{\omega_+^{\,2}-\frac{m^2}{4}}\right]&
\Bigg\}
 \end{split}
\end{equation}
\begin{equation}
 \begin{split}
A_4(p)= A_3(p) + \frac{\omega_+(p)}{2} \dq V(k) \Bigg\{ \Big[ \Cmp\Cm \pj  \hspace*{4cm}&\\ 
+\Smp\Sm \frac{J\pjp+ (J+1)\pjm}{2J+1} \Big] \frac{ A_4(q)}{\omega_+^{\,2}-\frac{m^2}{4}}&  \\
- \Smp \Cp\frac{\sqrt{J(J+1)}}{2J+1} \left[ \pjp - \pjm \right]\frac{ A_2(q)}{\omega_+^{\,2}-\frac{m^2}{4}}& \Bigg\}\,,
 \end{split}
\end{equation}
\normalsize
where $\omega_+(p) = \frac{1}{2} (\omega_1(p) + \omega_2(p))$ and $\varphi_\pm(p) = \frac{1}{2} (\varphi_1(p) \pm \varphi_2(p))$.

The equations above lead to the equations from section \ref{sec:BSE2m}, if we define a wave function as (see Appendix C)
\begin{equation}
 \psi_{i,\pm} (p) = \oh \left[ h_i(p) \pm \frac{m}{2}\left( 1 \pm \frac{m}{2 \,\omega_+(p)} \right) g_i(p) \right]\,,
\end{equation}
where
\begin{equation}\label{eq:onefiniteone}
 h_1(p) = \frac{A_1(p)}{\omega_+(p)}\,,
\end{equation}
\begin{equation}\label{eq:onefinitetwo}
 g_1(p) = \frac{A_2(p)}{\omega_+^{\,2}()-\frac{m^2}{4}}\,,
\end{equation}
\begin{equation}
 h_2(p) = \frac{A_3(p)}{\omega_+(p)}
\end{equation}
\begin{equation}
 g_2(p) = \frac{A_4(p)}{\omega_+^{\,2}(p)-\frac{m^2}{4}}\,.
\end{equation}
\normalsize

\section{Appendix C}

We derive the rest frame wave function and its projections on the
components propagating backward and forward in time \cite{G6}.

The wave function in the rest frame is
\begin{eqnarray}
\nonumber
  \psi_{JM}^{P}(m, \vec p) = \im  \int \frac{dp_0}{2 \pi} \, S_1 \left(p_0 + \frac{m}{2}, \vec p\right) 
  \chi_{JM}^P (m, \vec p\, )\, S_2 \left(p_0 - \frac{m}{2}, \vec p\right)\,.
\end{eqnarray}
The quark propagator can be splitted into
\begin{equation}
  S(p_0, \vec p) = S_+(p_0, \vec p) + S_-(p_0, \vec p)\,,
\end{equation}
where
\begin{equation}
 S_\pm (p_0, \vec p) = \im \frac{T_p P_\pm T_p^{\dagger}\gamma_0}{p_0 \mp \omega(p)\pm \im\epsilon}\,,
\end{equation}
\begin{equation}
  T_p = \exp\left[-\frac{1}{2}\vec\gamma\cdot\hat p \left(\frac{\pi}{2} - \varphi_p\right)\right]
\end{equation}
and projectors are
\begin{equation}
  P_\pm = \frac{1 \pm \gamma_0}{2}\,.
\end{equation}
The integration over $p_0$ may be performed with the help of equalities
$P_\pm \gamma_0 = \pm P_\pm$, $T_p^{\dagger} \gamma_0 = \gamma_0 T_p$ and leads to
\begin{eqnarray}
\nonumber
  \psi_{JM}^{P}(m, \vec p) = \frac{T_{p,1} P_+ T_{p,1} \chi_{JM}^{P}(m, \vec p) T_{p,2} P_- T_{p,2}}{2\omega_+(p) - m}\\
+\frac{T_{p,1} P_- T_{p,1} \chi_{JM}^{P}(m, \vec p) T_{p,2} P_- T_{p,2}}{2\omega_+(p) + m}\,.
\end{eqnarray}
Since $T_p^\dagger = T_p^{-1}$, the Foldy transformation of the wave function 
$\tilde{\psi}_{JM}^{P}(m, \vec p) = T_{p,1}^\dagger \psi_{JM}^{P}(m, \vec p) T_{p,2}^\dagger$ is
\begin{equation}
  \tilde{\psi}_{JM}^{P}(m, \vec p) = P_+ \psi_{+JM}^{P}(m, \vec p) P_- + P_- \psi_{-JM}^{P}(m, \vec p) P_+\,,
\end{equation}
where '$+$' and '$-$' components correspond to the propagation forward and backward respectively.
The result for the forward and backward propagating wave function for the category A is
\begin{equation}
  \psi_{\pm JM}^{P}(m, \vec p) \,=\,\, \psi_{1,\pm} (p)\, \gamma_5 Y_J(\hat p) \,\,\pm \,\, \psi_{2,\pm} (p)\left\{Y_J(\hat p) \times \vec \gamma \right\}_{JM}\,.
\end{equation}
For the category B the wave functions obey
\begin{equation}
\begin{split}
 \psi_{\pm JM}^{P}(m, \vec p) = \,\,\,
&\bigg[ \sqrt{\frac{J}{2J+1}} \psi_{1,\pm} (p) + \sqrt{\frac{J+1}{2J+1}} \psi_{2,\pm} (p) 
 \bigg] \left\{ Y_{J+1}(\hat p) \times \vec \gamma \right\}_{JM}\\
 + &\bigg[\sqrt{\frac{J+1}{2J+1}} \psi_{1,\pm} (p) - \sqrt{\frac{J}{2J+1}} \psi_{2,\pm} (p)
 \bigg] \left\{ Y_{J-1}(\hat p) \times \vec \gamma \right\}_{JM}\, .
 \end{split}
\end{equation}
The wave function components for both categories $\psi_{i,\pm} (p)$ are defined as
\begin{equation}
 \psi_{i,\pm} (p) = \oh \left[ h_i(p) \pm \frac{m}{2}\left( 1 \pm \frac{m}{2 \,\omega_+(p)} \right) g_i(p) \right]\, .
\end{equation}
% The wave functions for the heavy-light meson can be calculated using
% \begin{equation}
%  T_{p,2} =     \mathbb{1}
% \end{equation}
% for the heavy quark. 

\section{Appendix D}

\begin{figure}[ht]
 \centering
   \includegraphics[width=3.5in]{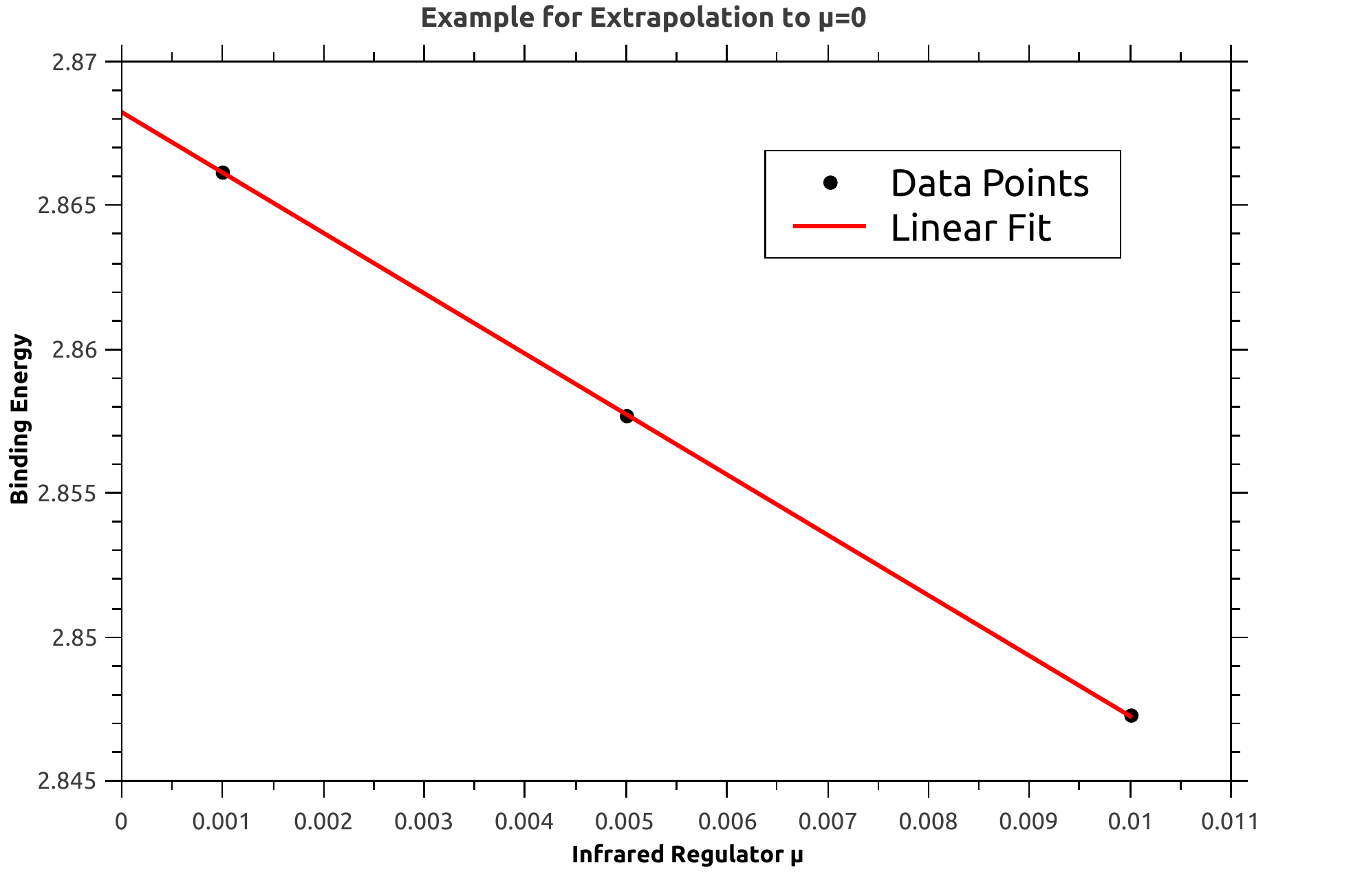}
   \rule{30em }{0.5pt}
   \caption{Calculated data points, linear fit}
   \label{fig:extrapol}
\end{figure}

Here we present a procedure for the numerical solution of the heavy-light BSE.
The binding energy spectrum and meson wave functions were calculated for three different 
values of the infrared regulator $\mu_{IR} = 0.01\sqrt\sigma$, $\mu_{IR} = 0.005\sqrt\sigma$ and $\mu_{IR} = 0.001\sqrt\sigma$.
The final results were obtained by extrapolation to the infrared limit $\mu_{IR} \rightarrow 0$, see Fig. \ref{fig:extrapol}.

The chiral angle and the single quark energy, 
being an input to the BSE, are obtained from iterative solution of the gap equation.

Taking into account the diagonalization (section \ref{sec:BSEhl}), the Bethe-Salpeter equation for an arbitrary $J$
may be viewed as
\begin{equation}
  (H(p\,|\,\omega(p)) + \epsilon) \psi(p) = \int dq~F(p, q\,|\, \varphi(p), \varphi(q))\psi(q)\,,
\label{inteq}
\end{equation}
where $\varphi(p)$ is the chiral angle, $\omega(p)$ is the single quark energy and $\psi(p)$ is the wave function.
We solve equations of this kind by expanding the unknown wave function in the basis
\begin{equation}
  \psi(p) = \sum_{i = 1}^{N} C_i\, \xi_i(p).
\end{equation}
To match the appropriate boundary  conditions, we choose 
\begin{equation}
  \xi_i(p) = p^{J} \exp(-\alpha_i p^2)
\end{equation}
for the mesons of the category A and
\begin{equation}
  \xi_i(p) = p^{|J - 1|} \exp(-\alpha_i p^2)
\end{equation}
for the mesons of the category B.
It is enough to use a relatively small number of 
Gaussians for a sufficient accuracy of the expansion.
Then the truncated equation (\ref{inteq}) becomes a system of linear equations
\begin{equation}
  (H(p\,|\,\omega(p)) + \epsilon) \sum_{i = 1}^{N} C_i\, \xi_i(p) = \int dq~F(p, q\,|\, \varphi(p), \varphi(q))\sum_{i = 1}^{N} C_i\, \xi_i(q)\,.
\label{syseq}
\end{equation}
Multiplying (\ref{syseq}) by $\xi_j(p)$ and integrating over $p$ we end up with the
generalized eigenvalue problem
\begin{equation}
  \epsilon\,D\,\vec{C} = (A + B)\, \vec{C}\,
\label{eigprob}
\end{equation}
where
\begin{eqnarray}
\nonumber
D_{ij} = \int dp\,\xi_i(p)\xi_j(p)\,,\\
\nonumber
A_{ij} = \int dp\,\xi_i(p)\xi_j(p)H(p\,|\,\omega(p))\,,\\
B_{ij} = \int dp\,\int dq\,\xi_i(p)\xi_j(q) F(p, q\,|\, \varphi(p), \varphi(q))\,.
\end{eqnarray}
The solution of the problem (\ref{eigprob}) leads to the spectrum of binding energies and to the corresponding 
meson wave functions.

\label{Bibliography}
% \lhead{\emph{Bibliography}} % Change the page header to say "Bibliography"
% \bibliographystyle{thesis} %\bibliographystyle{unsrtnat} % Use the "unsrtnat" BibTeX style for formatting the Bibliography
% \bibliography{Bibliography} % The references (bibliography) information are stored in the file named "Bibliography.bib"
% 
\bibliographystyle{unsrt}
\bibliography{bibliography}

\end{document}